\documentclass{jfm}

\usepackage{graphicx}
\usepackage{natbib}
\usepackage{subfig}
\usepackage[amssymb]{SIunits}
\usepackage{amsmath}
\usepackage{color}
\usepackage{epstopdf}

\ifCUPmtlplainloaded \else
  \checkfont{eurm10}
  \iffontfound
    \IfFileExists{upmath.sty}
      {\typeout{^^JFound AMS Euler Roman fonts on the system,
                   using the 'upmath' package.^^J}%
       \usepackage{upmath}}
      {\typeout{^^JFound AMS Euler Roman fonts on the system, but you
                   dont seem to have the}%
       \typeout{'upmath' package installed. JFM.cls can take advantage
                 of these fonts,^^Jif you use 'upmath' package.^^J}%
      }
  \else
  \fi
\fi

\ifCUPmtlplainloaded \else
  \checkfont{msam10}
  \iffontfound
    \IfFileExists{amssymb.sty}
      {\typeout{^^JFound AMS Symbol fonts on the system, using the
                'amssymb' package.^^J}%
       \usepackage{amssymb}%

      }{}
  \fi
\fi

\ifCUPmtlplainloaded \else
  \IfFileExists{amsbsy.sty}
    {\typeout{^^JFound the 'amsbsy' package on the system, using it.^^J}%
     \usepackage{amsbsy}}
    {}
\fi

\newcommand{\nom}{Nu_\omega}

\newsavebox{\astrutbox}
\sbox{\astrutbox}{\rule[-5pt]{0pt}{20pt}}

\title[The near-wall region of highly turbulent Taylor-Couette flow]{The near-wall region of highly turbulent Taylor-Couette flow}

\author[R. Ostilla M\'onico and others]%
{R\ls O\ls D\ls O\ls L\ls F\ls O\ns O\ls S\ls T\ls I\ls L\ls L\ls A - M\ls\'O\ls N\ls I\ls C\ls O$^1$,\break%
R\ls O\ls B\ls E\ls R\ls T\ls O\ns V\ls E\ls R\ls Z\ls I\ls C\ls C\ls O$^{2,1}$,\break%
S\ls I\ls E\ls G\ls F\ls R\ls I\ls E\ls D\ns G\ls R\ls O\ls S\ls S\ls M\ls A\ls N\ls N$^{3}$,\ns%
\and D\ls E\ls T\ls L\ls E\ls F\ns L\ls O\ls H\ls S\ls E$^1$}

\affiliation{$^1$Physics of Fluids, Mesa+ Institute, University of Twente, P.O. Box 217, 7500 AE Enschede, The Netherlands\\[\affilskip]
$^2$Dipartimento di Ingegneria Industriale, University of Rome ``Tor Vergata'', Via del Politecnico 1, Roma 00133, Italy\\[\affilskip]
$^3$Department of Physics, University of Marburg, Renthof 6, D-35032 Marburg, Germany}

\date{\today}

\begin{document}
 
\maketitle

\begin{abstract}
Direct numerical simulations of the Taylor-Couette (TC) problem, the flow between two coaxial and independently rotating cylinders, have been performed. The study focuses on TC flow with mild curvature (small gap) with a radius ratio of $\eta=r_i/r_o=0.909$, an aspect ratio of $\Gamma=L/d=2\pi/3$, and a stationary outer cylinder. Three inner cylinder Reynolds of $1\cdot10^5$, $2\cdot10^5$ and $3\cdot 10^5$ were simulated, corresponding to frictional Reynolds numbers between $Re_\tau\approx 1400$ and $Re_\tau \approx 4000$. An additional case with a large gap, $\eta=0.5$ and driving of $Re=2\cdot10^5$ was also performed. Small-gap TC was found to be dominated by spatially-fixed large-scale structures, known as Taylor rolls (TRs). TRs are attached to the boundary layer, and are active, i.e. they transport angular velocity through Reynolds stresses. An additional simulation with inner cylinder Reynolds number of $Re=1\cdot10^5$ and fixed outer cylinder with an externally imposed axial flow of comparable strength as the wind of the TRs was also conducted. The axial flow was found to convect the TRs without any weakening effect. For small-gap TC, evidence for the existence of logarithmic velocity fluctuations, and of an overlap layer, in which the velocity fluctuations collapse in outer units, was found. Profiles consistent with a logarithmic dependence were also found for the angular velocity in large-gap TC, albeit in a very reduced range of scales. Finally, the behaviour of both small- and large-gap TC was compared to other canonical flows. Small-gap TC has similar behaviour in the near-wall region to other canonical flows, while large-gap TC displays very different behaviour.
\end{abstract}


\section{Introduction}

Direct numerical simulations (DNS) are a very versatile tool for the study of turbulence, and have led to deep insight on its nature in the last 30 years. The seminal work of \cite{kim87} began a line of simulations of wall-bounded flows which is very active up to present days. \cite{kim87} studied a particular kind of wall-bounded flow, namely pressure-driven (Poiseulle) flow bounded by two parallel plates, from now on referred to as channel flow.  After this seminal work, which achieved $Re_\tau=180$, simulations of wall-bounded flows focused mainly on such channel flows, even though some attention was given to zero-pressure-gradient boundary layers (ZPGBL) and pipes. Present day simulations achieve much higher frictional Reynolds numbers: for channels $Re_\tau=4600$ \citep{ber14,loz14} and $Re_\tau\approx5200$ even more recently \citep{lee15}, $Re_\tau=2000$ for ZPGBL \citep{sil13} and $Re_\tau\approx1100$ for pipes \citep{wu08}. DNS has allowed for detailed studies of the near-wall energy cascade and for correlations between different canonical flows to be established. This has led to further understanding of the attached-eddy model of the logarithmic layer by \cite{tow76}, further developed by \cite{per82,per86}. We refer the reader to \cite{jim12} for a recent review of advances in this field.
 
DNS of \emph{shear} driven flow between two parallel plates, i.e. plane Couette (PC) flow, has been even more challenging. This is due to the extremely large and wide structures present in the turbulent flow, seen both experimentally and numerically \citep{bec95}. DNS of PC requires much larger computational boxes than that of channels, a factor ten in both spanwise and streamwise directions, because of the substantailly longer correlation lengths \citep{tsu06}. This meant that only recently $Re_\tau=550$ was achieved, namely by \cite{avs14} and more recently $Re_\tau\approx1000$ by \cite{pir14}. An alternative to having two independently moving parallel plates is having two independently \emph{rotating coaxial cylinders}. Such system is known as cylindrical Couette flow or Taylor-Couette (TC) flow. TC is a closed system, which makes experimental realizations easier to construct. Furthermore, TC does not require the large computational boxes of PC, which makes higher $Re_\tau$ easier to achieve in DNS. For a recent summary of the state-of-the-art of high Reynolds number Taylor-Couette, we refer the reader to the review by \cite{gro16}.

However, TC flow is fundamentally different from PC flow. Unlike PC (and all the flows mentioned previously) TC is linearly unstable if $(d(r^2\omega)/dr)^2 < 0$, where $r$ is the radial coordinate and $\omega$ the angular velocity. \cite{fai00} explored the transition from PC to TC, and found that only for radius ratios $\eta=r_i/r_o>0.99$, where $r_i$ and $r_o$ are the inner and outer cylinder radii respectively, the sub-critical PC instabilities overcame the super-critical TC instabilities. This linear instability causes the formation of large-scale structures, called Taylor vortices after the seminal work by \cite{tay23}, which become the defining feature of the flow. With increasing driving, these structures feature transitions from laminar Taylor vortices, to wavy Taylor vortices and finally to turbulent Taylor vortices. Taylor rolls have been observed in both experiments and simulations, even up to $Re=10^6$ \citep{hui14}. In the turbulent regime, these differences may be smaller. \cite{bra15} analysed the transport of momentum in both TC and rotating PC flow for the limit of vanishing curvature, finding a smooth transition in many of the flow quantities including mean streamwise velocity profiles, especially for the case of co-rotating cylinders.

\cite{bra73} was the first to note ``the surprisingly large effect exerted on shear-flow turbulence by curvature of the streamlines in the plane of the mean shear'', when studying boundary layers over curved surfaces. For curved channels \cite{hun79} and \cite{hof85} found that adding a very weak destabilizing (concave) curvature $L_\nu/R\sim0.01$, where $L_\nu$ is the boundary layer thickness and $R$ the radius of curvature, produced significant differences in the mean and fluctuation velocity profiles. This is because destabilizing curvature adds a new mode of instability, which is reflected as Taylor-G\"{o}rtler vortices in the flow and this leads to a change in the flow dynamics. We note that pipes are not a member of this group. The natural curvature of pipes is in planes perpendicular to the mean shear, so its effects are milder. 

Both channel flow and TC have been used as a playground for investigating drag reduction through deformable bubbles or riblets \citep{cho93,ber05,lu05,gil13,zhu15}. From the previous discussion, we would expect significant differences between both systems, but instead, very similar results were obtained. In both cases drag was reduced, and both the bubbles and riblets modified the boundary layer. As such similar drag reduction mechanisms can be seen in both systems, this may be taken as an indication that  the boundary layers in both systems may also be similar. Further evidence for such a universal behaviour was provided by recent experiments: \cite{hui13} measured the mean velocity profiles in TC and obtained a von Karman constant $\kappa\approx 0.4$ for the highest Reynolds numbers achieved (with $Re_\tau\approx30000$), in line with what is seen in experimental pipes \citep{bai14}. Outside the boundary layer, things change. The Taylor vortices effectively redistribute angular momentum, leading to a constant angular momentum between the cylinders \citep{wer99b}. If the Reynolds number is large enough, the wavelength of the Taylor rolls play very little effect in determining the torque required to drive the cylinders \citep{ost14e,ost15}. However, their signature in the velocity field is still apparent, clearly appearing in the mean fields. An analogous structure which fills the entire domain and is persistent in time was also found in curved channels with fully developed turbulence: large scale Taylor-G\"{o}rtler vortices \citep{hun79,mos86}.

However, many fundamental differences seem to exist between pipe \& channel flows and TC flow. To explore them, in this manuscript we further characterize the TC system at high Reynolds numbers in general, and in particular the effect of the Taylor rolls.  To do so, five high Reynolds number DNSs of TC flow were performed. All simulations were done for pure inner cylinder rotation, with a stationary outer cylinder and with an axial periodicity aspect ratio $\Gamma=L_z/d=2\pi/3$, where $L_z$ is the axial periodicity length. With this $\Gamma$, the system fits a single Taylor roll pair of wavelength $\lambda_{TR}=2\pi/3$, which is axially fixed. We focus on $\eta=0.909$ which has a mild enough curvature to make it linearly unstable, but small enough not to be dominated by the streamline topology. Three simulations at shear Reynolds numbers of $Re_s=d\omega_i/\nu=10^5$, $Re_s=2\cdot 10^5$ and $Re_s=3\cdot 10^5$, were performed, where $d$ is the gap-width $d=r_o-r_i$, $\omega_i$ and $\omega_o$ are the inner and outer cylinder angular velocity respectively, and $\nu$ is the kinematic viscosity of the fluid. This results in frictional Reynolds numbers at the inner cylinder between $Re_{\tau,i}\sim1000$ and $Re_{\tau,i}\sim4000$, with the inner cylinder frictional Reynolds number $Re_{\tau,i}$ defined as $Re_{\tau,i}=u_{\tau,i} d/(2\nu)$, where the frictional velocity is $u_{\tau,i}=\sqrt{\tau_{w,i}/\rho}$, with $\rho$ the fluid density and $\tau_w$ the stress at the inner cylinder wall. 

To further probe the effect of the Taylor rolls, two additional simulations were run. The first was a simulation of large-gap, curvature-dominated TC, with a radius ratio of $\eta=0.5$, an aspect ratio $\Gamma=2\pi/3$ and a driving of $Re_s=2\cdot 10^5$ which resulted in a frictional Reynolds number at the inner cylinder of $Re_{\tau,i}\approx 3100$. This simulation is in the region of the TC parameter space where Taylor rolls are heavily weakened or have even disappeared completely \citep{ost14e}. Furthermore, to probe the spatial stability of the rolls to a transversal velocity, an additional simulation at $\eta=0.909$ and $Re_s=10^5$ with an imposed axial (spanwise) flow was performed. This axial flow, generated by a pressure gradient, delays the onset of turbulence, but for high enough driving it does not impede the linear instability of the flow \citep{cha60a,cha60b}. Beyond the onset of turbulence and for weak axial flows, the toroidal rolls are simply convected upwards. With increasing axial flow, spiral rolls are formed. These were first observed experimentally by \cite{sny62}, and subsequently by many others, both experimentally and numerically \citep{sch64,tak81,ng82,tsa94}. For a comprehensive overview of the different flow regimes for low Reynolds number TC flow with rotating inner cylinder and an axial pressure gradient, we refer the reader to \cite{lue92,wer99b,hwa04}. In these cases, the flow behaves mainly as a linear superposition of the imposed axial flow and the TC rolls. Here, we focus on the large Reynolds number case, and non-linear effects may, or may not arise. The imposed axial velocity had $U_w \approx \frac{1}{10} U_i$, where $U_i$ is the inner cylinder velocity $U_i=r_i\omega_i$. This $U_w$ is of the order of magnitude of the characteristic velocity of the Taylor rolls, but still small enough such that the system dynamics is not dominated by this secondary flow. The pressure gradient generating this flow was spatially uniform, but variable in time. Its magnitude was controlled by a proportional-integral (PI) controller which acted on the difference between the desired axial flow rate and the actual axial flow rate. 

The manuscript is organized as follows. $\S$\ref{sec:numset} presents the numerical setup and details of the simulations. $\S$\ref{sec:recs} presents the results and discussions. The findings are summarized in the final section $\S$\ref{sec:summ}, where also an outlook on future work is given. 

\section{Numerical details}
\label{sec:numset}

The DNS were performed using a second order centred finite difference scheme with fractional-time stepping  \citep{ver96,poe15}. This scheme has been used and validated extensively in the context of TC (cf. comparison to experiments in \cite{ost14,ost14e}). In order to perform the simulations at high Reynolds numbers ``small'' computational boxes were used. Instead of simulating the full azimuthal extent of the cylinder, a cylindrical wedge was simulated by imposing a rotational symmetry $n_{sym}$ of order $n_{sym}=20$ for $\eta=0.909$. This gives an azimuthal extent at the mid-gap of $1.05\pi d$. For $\eta=0.5$, $n_{sym}=3$ was used, giving an azimuthal extent of $\pi d$ at the mid-gap. The axial box size was also minimal, $\Gamma$ was set to $\Gamma=2\pi/3$ for all simulations, meaning that a single Taylor roll pair of wavelength $\lambda_{TR}=2.09$ could fit in to the domain. Simulating more than one roll-pair was shown to be unnecessary for producing accurate results for the torque by \cite{bra13}. Furthermore, very sharp dropoffs of the axial spectra can be seen in \cite{don07} for wavelengths larger than one roll.

\begin{table}
  \begin{center} 
  \def~{\hphantom{0}}
  \begin{tabular}{|c|c|c|c|c|c|c|c|c|c|c|}
  \hline
  Case & $Re_s$ & $N_\theta$ & $N_r$ & $N_z$ & $\Delta x^+$ & $\Delta z^+$ & $\nom$ & $Re_{\tau,i}$ \\ 
  \hline
  R1 & $1\cdot10^5$ & $1024$ & $1024$ & $2048$ & $9.1$ & $2.7$ & $69.5\pm0.2$ & $1410$ \\
  R2 & $2\cdot10^5$ & $1536$ & $1536$ & $3072$ & $11.4$ & $3.4$ & $126\pm2.1$ & $2660$ \\
  R3 & $3\cdot10^5$ & $2048$ & $1536$ & $3072$ & $12.6$ & $5.1$ & $171\pm2.5$ & $3920$ \\
  AF & $1\cdot10^5$ & $1024$ & $1024$ & $2048$ & $9.1$ & $2.7$ & $66.2\pm0.5$ & $1390$ \\
  E5 & $2\cdot10^5$ & $1024$ & $1536$ & $3072$ & $12.5$ & $4.0$ & $77.2\pm0.5$ & $3080$ \\
  \hline
 \end{tabular}
 \caption{Details of the numerical simulations. The first column is the name with which the simulation will be refereed to in the manuscript. The second column is $Re_s$, the shear Reynolds number. The third to fifth columns represent the amount of points in the azimuthal, radial and axial directions, while the sixth and seventh columns show the resolution in inner wall-units at the mid-gap, $\Delta x^+=\frac{1}{2}(r_i+r_o)\Delta\theta^+$. The eighth column is the non-dimensional torque $\nom$. The ninth column is $Re_{\tau,i}=u_{\tau,i}d/(2\nu)$, the frictional Reynolds number at the inner cylinder. $Re_{\tau,o}$ can be obtained from $Re_{\tau,o}=\eta Re_{\tau,i}$. The AF case has an imposed axial flow with a mean velocity $U_w = \frac{1}{10}r_i\omega_i$, and the E5 case is the large-gap case with $\eta=0.5$.  }
 \label{tbl:reso}
\end{center}
\end{table}

Full details of the numerical simulations are presented in Table \ref{tbl:reso}. We mention that the choice of $\lambda_{TR}=2.09$ is arbitrary. Fixing the wavelength of the roll allows for comparison of fluctuations in the bulk across Reynolds numbers, as these depend on roll wavelength \citep{ost15}.  We also note that experiments have shown that states with a different amount of rolls, and thus a different roll wavelengths are accessible for exactly the same flow parameters \citep{mar14,hui14}. For large Reynolds number, these are generally rectangular vortices, with $\lambda_{TR}>2$. To demonstrate that our small boxes are sufficient, a more detailed discussion about the effect of the computational box-size, and of the effects of $\lambda_{TR}$ on TC flow we refer the reader to \cite{ost15}. There, it is shown that for $\eta=0.909$ computational boxes with $n_{sym}=20$ and $\Gamma=2.09$ are large enough for the autocorrelations to smoothly change sign in both axial and azimuthal direction, and in the latter case, to also approach zero. As previously mentioned, the azimuthal (streamwise) correlations decay much faster in TC flow than in PC flow \citep{tsu06}, and thus small boxes can be used. 

The axial and azimuthal spectra shown in section \ref{sub:spec} demonstrate that the mesh is sufficient to capture the small scales. We mention that $\Delta z^+ \approx 5$ is a marginally resolved case, while for the azimuthal direction, coarser resolutions of up to $r\Delta \theta^+\approx 12$ can be used without loss of accuracy or adding numerical dispersion which changes the physics of the problem.

\begin{figure}
  \centering
  \includegraphics[width=0.49\textwidth]{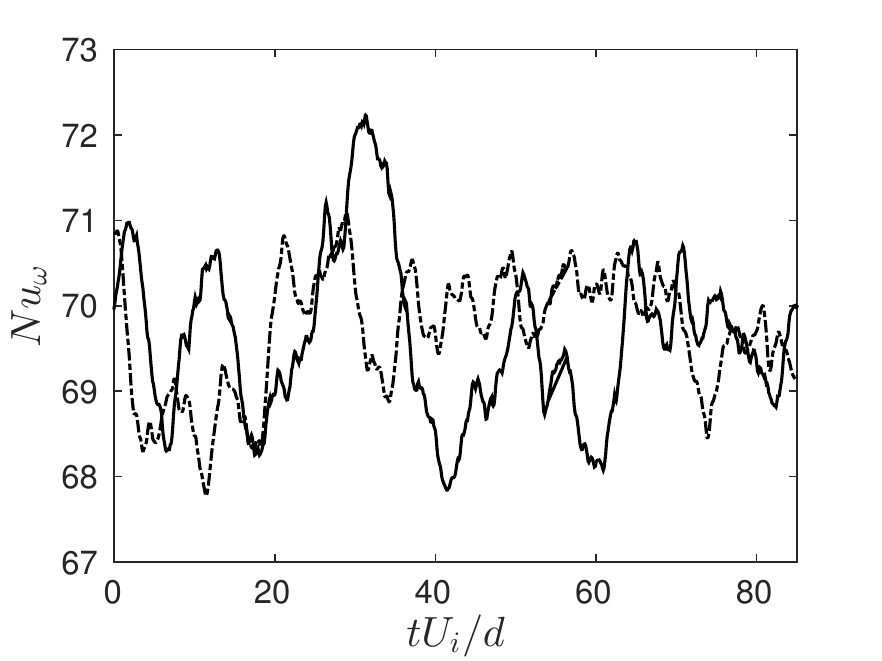}
  \includegraphics[width=0.49\textwidth]{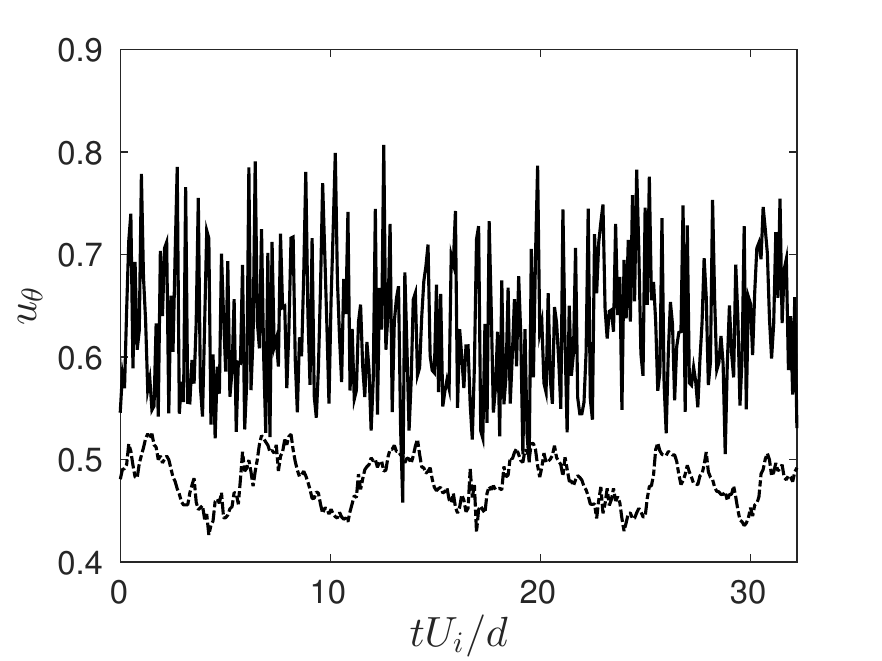}
  \caption{ Left: Instantaneous $\nom$ at inner (solid curve) and outer (dash-dot curve) cylinders for the R1 simulation. Right: instantaneous azimuthal velocity near the inner cylinder ($r^+\approx 12$, upper \& solid curve) and in the mid-gap (lower \&  dash-dot curve) for the R1 case. The temporal origin is arbitrary. Fluctuations are on timescales of the order $\mathcal{O}(d/U_i)$.  }
\label{fig:torquetime}
\end{figure}

The simulations were run for an additional $50$ large eddy turnover times based on $d/U_i$ after transient behaviour had died out. In time units based on the frictional velocity $u_\tau$ and half the gap width, this is between $3$ and $4$ turnover times. This might seem small when comparing to channels, but it is sufficient for TC flow with pure inner cylinder rotation. The characteristic time for TC appears to be $d/U_i$. \cite{don08} already showed that the decay of temporal autocorrelations in TC happens in $t\approx 3d/U_i$. This fast timescale is further quantified in the left panel of Figure \ref{fig:torquetime}, which presents the instantaneous non-dimensional torque $\nom=T/T_{pa}$, where $T$ is the torque and $T_{pa}$ the torque in the purely-azimuthal flow, for the R1 case and the right panel of Figure \ref{fig:torquetime}, which shows the instantaneous azimuthal velocity at two points for the R1 case. The fluctuation time-scale at the mid-gap is much faster than the one inside deep inside the boundary layer, probably due to the influence of the large-scale structures.

As we will see in more detail in $\S$\ref{sub:spec}, the Reynolds stresses which transport angular velocity are mainly localized in low-wavelength eddies, which have velocities of order $\mathcal{O}(U_i)$. This is not the case in channel flow, where the largest eddies are inactive, i.e. they do not transport Reynolds-stresses \citep{tow76,hoy06}. Reynolds-stresses are transported by boundary layer detachments, or streaks, with velocity scales of $\mathcal{O}(u_\tau)$, which then naturally leads to the timescales $\mathcal{O}(d/u_\tau)$ for shear transport. However, for the large scales to saturate in energy, much larger times are needed. Large scales are ``fed'' by detachments from the boundary layers with velocity $\mathcal{O}(u_\tau)$. This means that the \emph{transients} in TC flow have timescale $\mathcal{O}(d/u_\tau)$. Therefore, we define the non-dimensional time based on a large eddy turnover time as $\tilde{t}=tU_i/d$. 

In the following sections, we will use several kinds of averaging and quantification of fluctuations which are detailed here. The averaging operator of a field $\phi$ with respect to the independent variable $x_i$ is denoted $\langle \phi \rangle_{x_i}$. We thus define the two different averages: $\bar{\phi}=\langle \phi \rangle_{\theta,t,z}$, and 
$\hat{\phi}=\langle \phi \rangle_{\theta,t}$. The first type of averaging is analogous to that of homogeneous flows, while the second type of averaging is similar to the one used in curved channels by \cite{mos86}, where large-scale structures are also present. Using this, we can define two sorts of measures of the fluctuation level. First, is the ordinary definition for homogeneous flows: $\phi^\prime = \bar{(\phi^2)}-(\bar{\phi})^2$, and a second definition, used to separate the effect of the axial inhomogeniety due to the large-scale structure: $\phi^\star=\langle\hat{(\phi^2)}-(\hat{\phi})^2\rangle_z$. 

Inner cylinder wall units are defined using $u_{\tau,i}$ as a velocity scale and $\delta_{\nu,i}=\nu/u_{\tau,i}$ as a length scale. The mean azimuthal profile is defined as a velocity difference, i.e. $U^+ = (U_i-\bar{u}_\theta)/u_{\tau,i}$, where $\langle \phi \rangle_{x_i}$ denotes the variable $\phi$ averaged with respect to $x_i$. $r^+$ is the distance to the inner cylinder $r^+=(r-r_i)/\delta_{nu,i}$. Outer cylinder wall units are defined using $\delta_{\nu,o}$ and $u_{\tau,o}$, and $r^+=(r_o-r)/\delta_{\nu,o}$. The frictional Reynolds number at the outer cylinder is obtained from $Re_{\tau,o}\approx\eta Re_{\tau,i}$. The non-dimensional distance from the inner cylinder in outer variables is defined as $\tilde{r}=(r-r_i)/d$, and the angular momentum as $L=ru_\theta$.

\section{Results}
\label{sec:recs}

\subsection{Characterization of the large-scale structures}
\label{sub:flowvis}

\begin{figure}
  \centering
   \includegraphics[trim=2.5cm 0cm 3.5cm 0cm,clip=true,height=6cm,angle=-0]{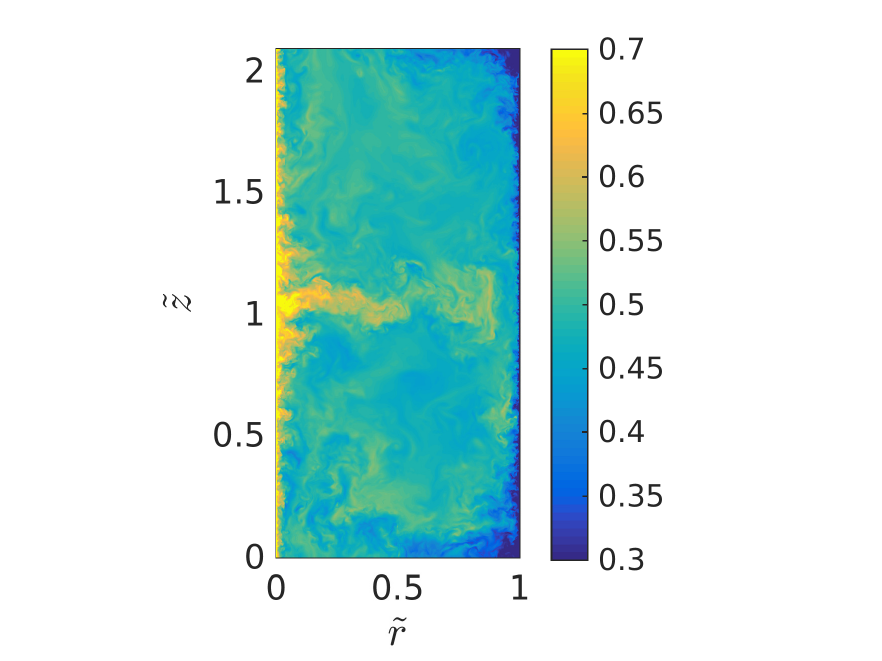} 
   \includegraphics[trim=3.5cm 0cm 3.5cm 0cm,clip=true,height=6cm,angle=-0]{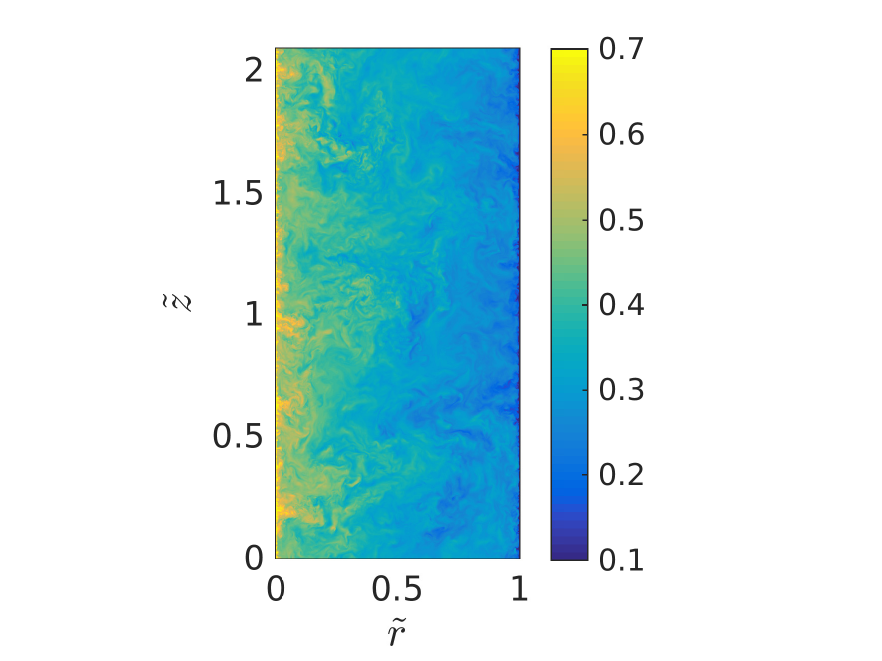}  
  \caption{ Visualization of the instantaneous azimuthal velocity $u_\theta$ for an azimuthal cut for the R2 (left) and E5 (right) cases. The large scale roll can be clearly seen in the left panel. These patterns are absent in the right panel, which shows a larger degree of homogeneity in the axial direction. }
\label{fig:q1instE5}
\end{figure}

For the R1-R3 cases, consisting of a radius ratio of $\eta=0.909$ and a stationary outer cylinder, large scale Taylor rolls exist which fill up the entire domain. Taylor rolls have a fixed axial position and are effectively axisymmetric, even if they are modulated by low-frequency azimuthal waves which do not cause the rolls to vary their spatial location or strength substantially \citep{ost15}. The rolls redistribute angular momentum across the gap, and imprint large scale patterns which are clearly visible when looking at the velocity field. The left panel of figure \ref{fig:q1instE5} shows a pseudocolour plot of the instantaneous azimuthal velocity for the R2 case. A segretation between regions from where herringbone-like streaks detach (eject) from the boundary layer, and regions where they impact on the boundary layer. We refer to the small-scale structures seen in TC flow as herringbone-like streaks, following \cite{don07}. These structures can also be thought of as plumes in the context of convection- as TC flow can be thought of a system in which angular momentum is transported from across the gap.

In figure \ref{fig:q1instE5} we can visualize a region of high azimuthal velocity, from where these herringbone streaks detach from the inner cylinder boundary layer around $\tilde{z}=1$. Alternatively in the region around $\tilde{z}=0$, streaks detach from the outer cylinder and impact the inner cylinder boundary layer. This detachment/impacting pattern contrasts with the large gap E5 case, shown in the right panel of the figure. Here, the large scale structures disappear completely, and the herringbone streaks detach and impact all over the cylinder wall. The lack of a large-scale structure was attributed to the asymmetry between the inner and the outer cylinder in \cite{ost14e}. It is worth noting that in some regions of the parameter space, i.e. for weakly counter-rotating cylinders, large scale structures can reappear again \citep{vee15}.

\begin{figure}
  \centering
  \includegraphics[trim=3cm 0cm 3cm 0cm,clip=true,height=6cm,angle=-0]{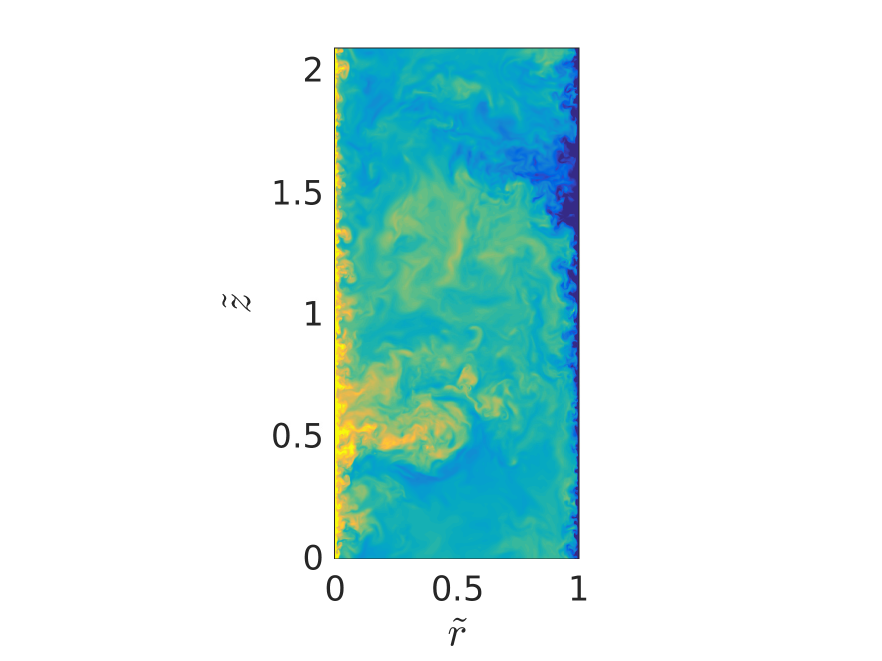}
  \includegraphics[trim=4cm 0cm 3cm 0cm,clip=true,height=6cm,angle=-0]{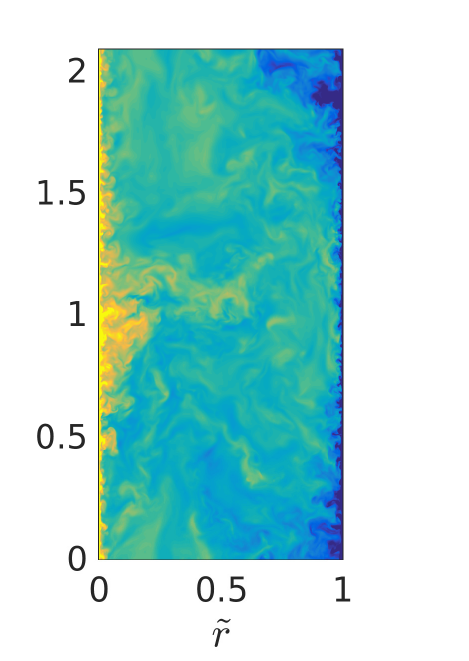}
  \includegraphics[trim=3.5cm 0cm 3.5cm 0cm,clip=true,height=6cm,angle=-0]{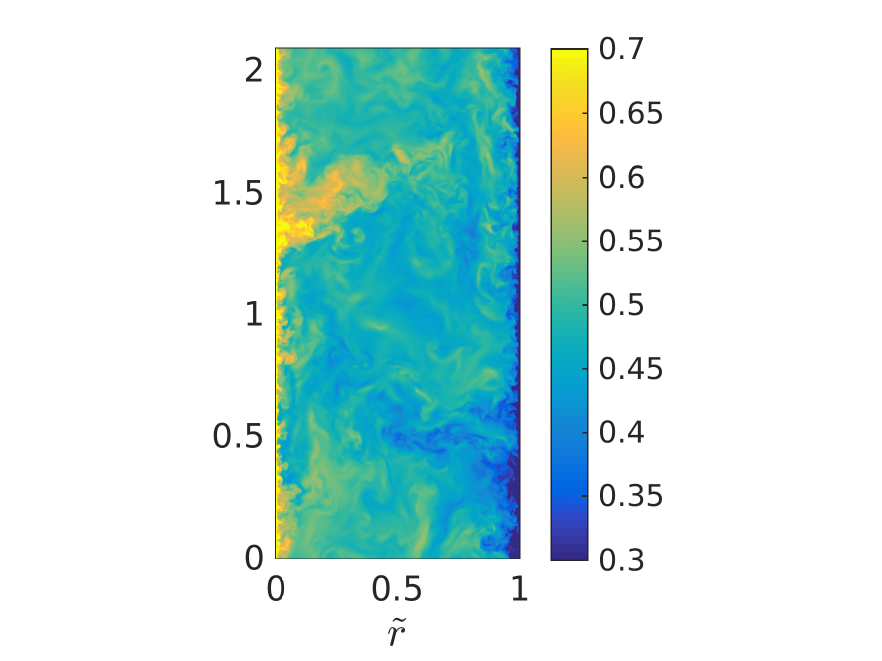}
 \caption{ Visualization of the instantaneous azimuthal velocity $u_\theta$ for an azimuthal cut for the AF case at times separated by $\tilde{t}=5$. The large scale pattern can be clearly seen all panels. The roll can be seen to slightly bend upwards due to the presence of an axial mean flow. }
\label{fig:q1inst}
\end{figure}

We attempted to unfix the rolls by the addition of an axial pressure gradient, which sustains a mean axial flow $U_w=\frac{1}{10}r_i\omega_i$  (case AF in Table \ref{tbl:reso}). Because this flow is strong compared to that induced by the Taylor rolls they do not switch from toroidal rolls to spiral rolls. However, they are no longer fixed in space, and instead they are convected upwards. Thus, the flow becomes again axially homogeneous in a statistical sense. Figure \ref{fig:q1inst} shows a pseudocolour plot of the instantaneous azimuthal velocity for the AF case in three separate instances in time. The large scale patterns caused by the presence of the underlying roll can be appreciated in all the panels. The (weak) mean axial velocity  does not prevent the formation of the Taylor rolls. Instead, as the velocity is ``small'', the rolls are slowly convected upwards in the computational domain, reappearing on the other side due to the axial periodicity. This is the same as what was seen for the low Reynolds number studies of TC flow with an axial pressure gradient, where the Taylor rolls still formed and the axial flow was not large enough to trigger the formation of spirals. The inherent linear instability of the system, present even with an axial flow \citep{cha60a,cha60b}, causes the formation of these large structures even at high Reynolds numbers.

\begin{figure}
  \centering
  \includegraphics[width=0.49\textwidth]{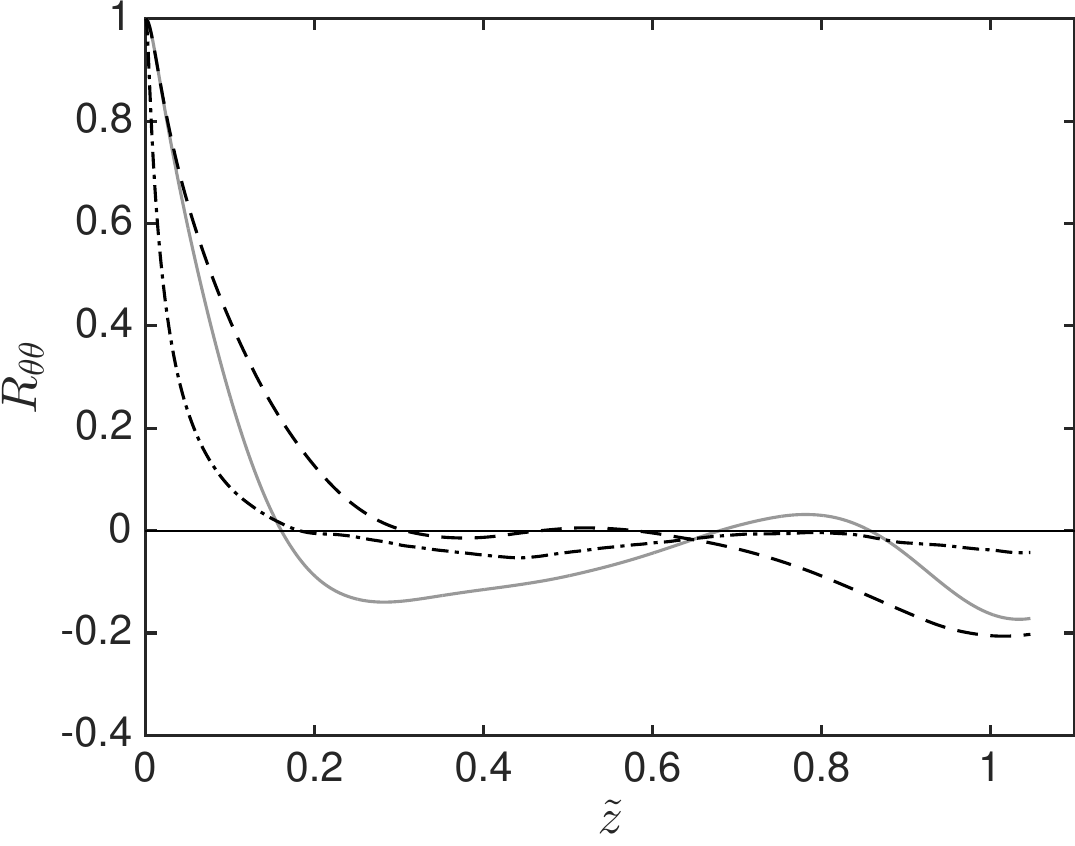}
  \includegraphics[width=0.49\textwidth]{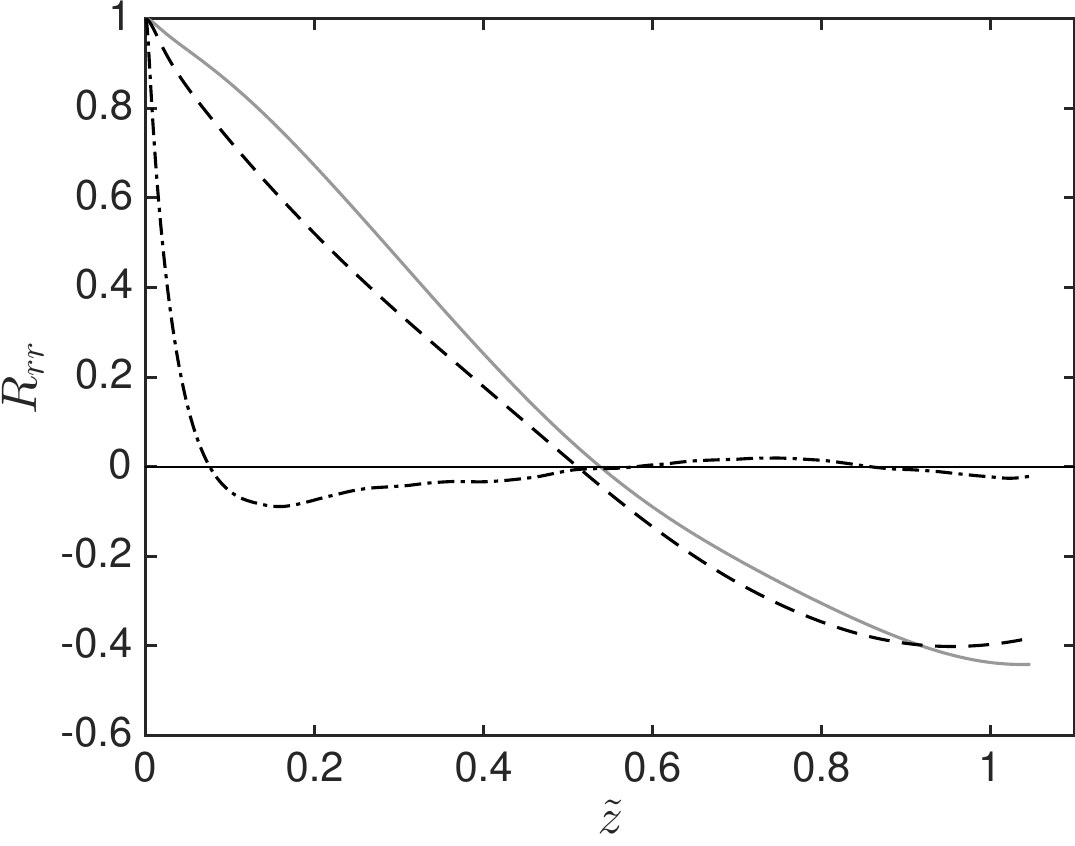}\\
  \caption{ Autocorrelation functions for the azimuthal and radial velocities in the axial directions at the mid-gap $\tilde{r}=0.5$. The signature of the Taylor rolls can be seen on the axial autocorrelations for the R1 and AF case, which are however absent for the E5 case. }
\label{fig:autocorrs}
\end{figure}

To further quantify the disappearance of the large-scales, we show in figure \ref{fig:autocorrs} the autocorrelations of the azimuthal and radial velocities at the mid-gap in the axial direction for the R1, AF and E5 cases. The signature of the rolls, as a large negative autocorrelation at half the roll wavelength, is clearly seen for the R1 and AF cases, both at small gaps with $\eta=0.909$ and it is absent from the axial autocorrelations of the large-gap E5 case. Strangely, an additional minimum for the azimuthal velocity is shown at $\tilde{z}=0.25$, which was seen already in \cite{ost15} consistently at an eighth of the vortex wavelength for different computational domain sizes.

\begin{figure}
  \centering
  \includegraphics[width=0.49\textwidth]{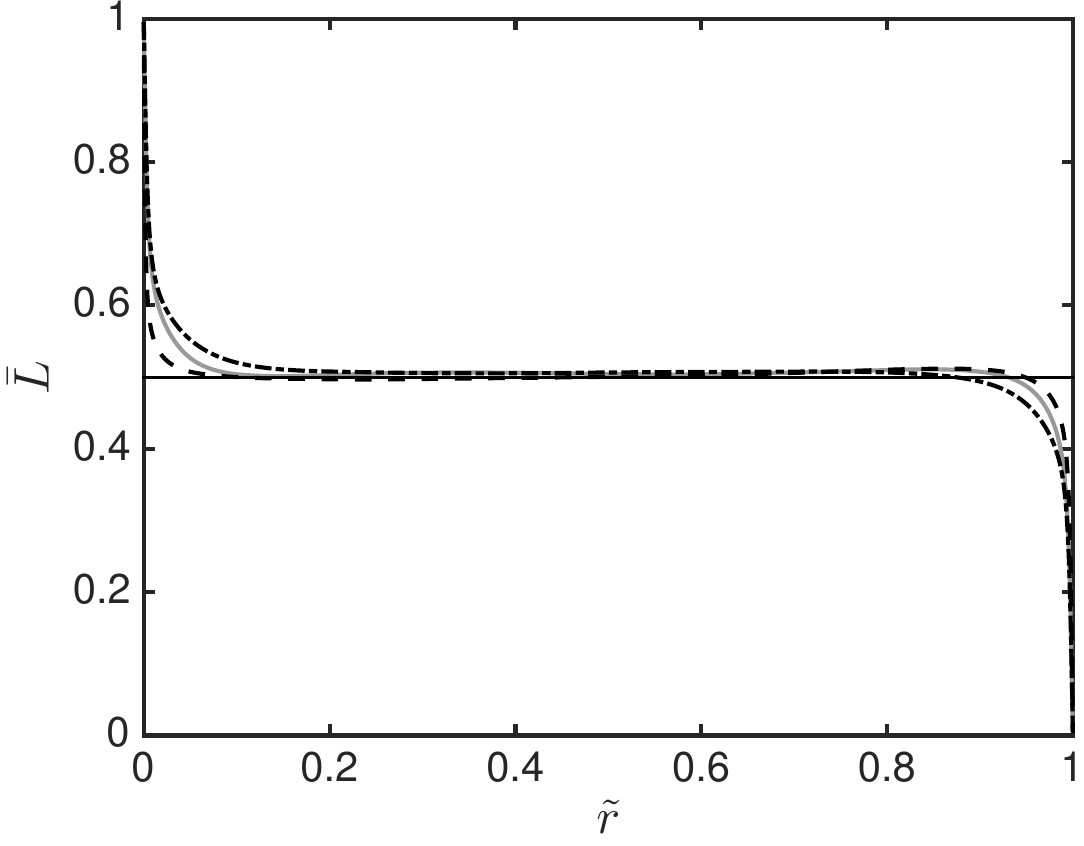}\\
  \caption{ Angular momentum $\bar{L}$ for the R1, AF and E5 cases. A region of constant angular momentum $\bar{L}\approx0.5$ (indicated in the panel by a vertical line) can be seen in the ``bulk''. Symbols: R1 light gray solid curve, E5 black dashed curve, AF black dash-dot curve.}
\label{fig:angmom}
\end{figure}

Figure \ref{fig:angmom} shows the averaged angular momentum $\bar{L}$ for the E5, AF and R1 cases. Angular momentum is practically constant at the centre of the domain, and equal to the mean angular momentum of both cylinders for all cases. This effect is similar to what is seen at low Reynolds numbers \citep{mar84b,wer94b}, and has also recently been observed by \cite{bra15} at high Reynolds numbers for several radius ratios and rotation ratios. From this figure, we can clearly distinguish the bulk region, which has constant angular momentum, and the thin boundary layers, where angular momentum has very steep gradients. We note that even if the large-scale structures disappear in the E5 case, angular momentum is still effectively redistributed by turbulence across the entire gap.

\subsection{Mean streamwise profiles}
\label{sec:onepoint}

\begin{figure}
  \centering
  \includegraphics[width=0.49\textwidth]{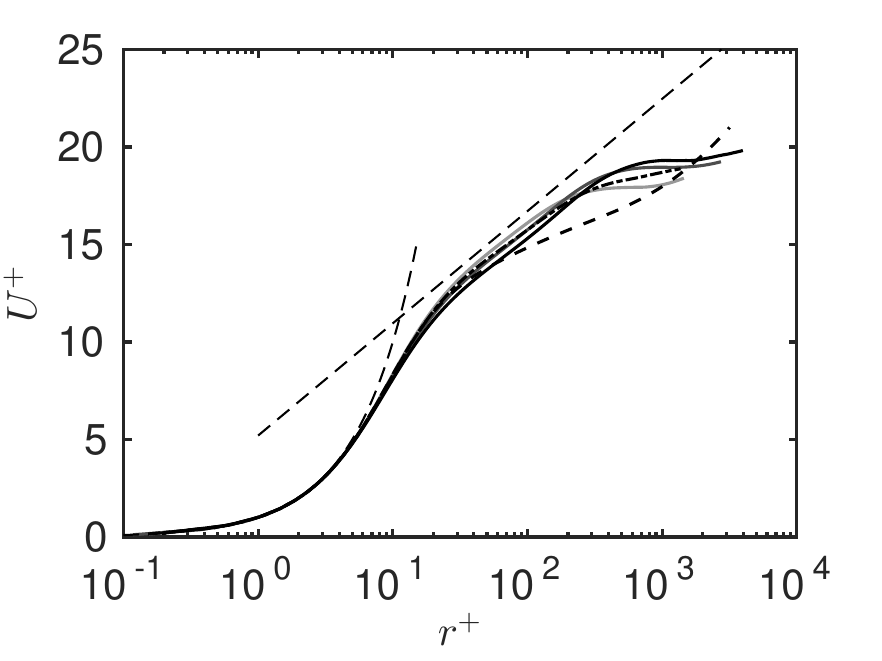}
  \includegraphics[width=0.49\textwidth]{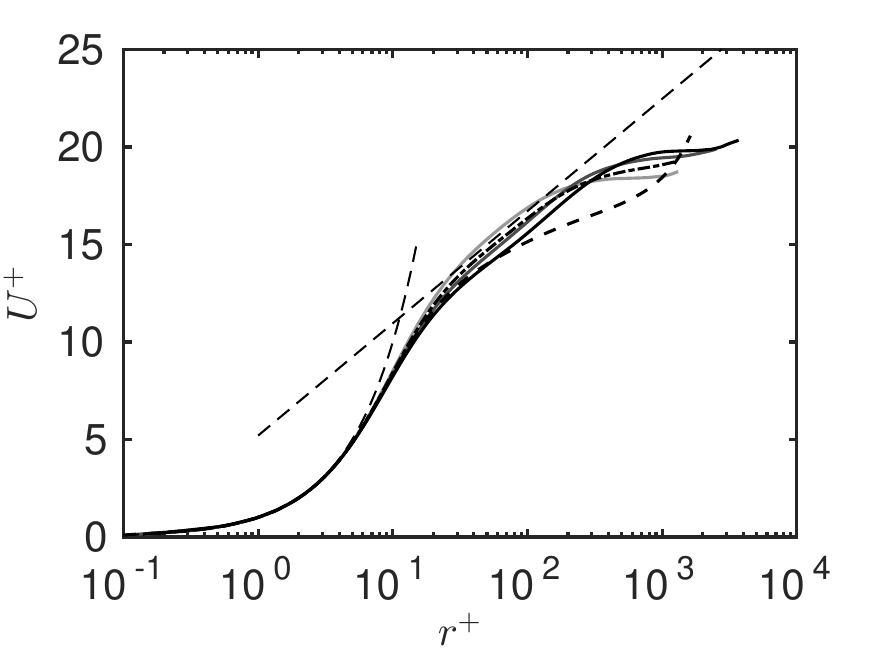}\\
  \includegraphics[width=0.49\textwidth]{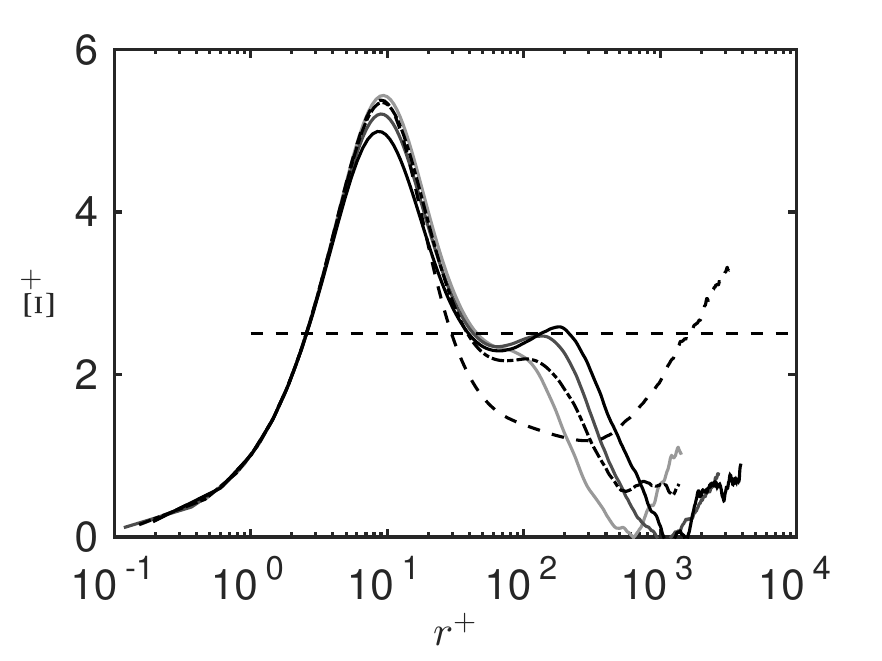}
  \includegraphics[width=0.49\textwidth]{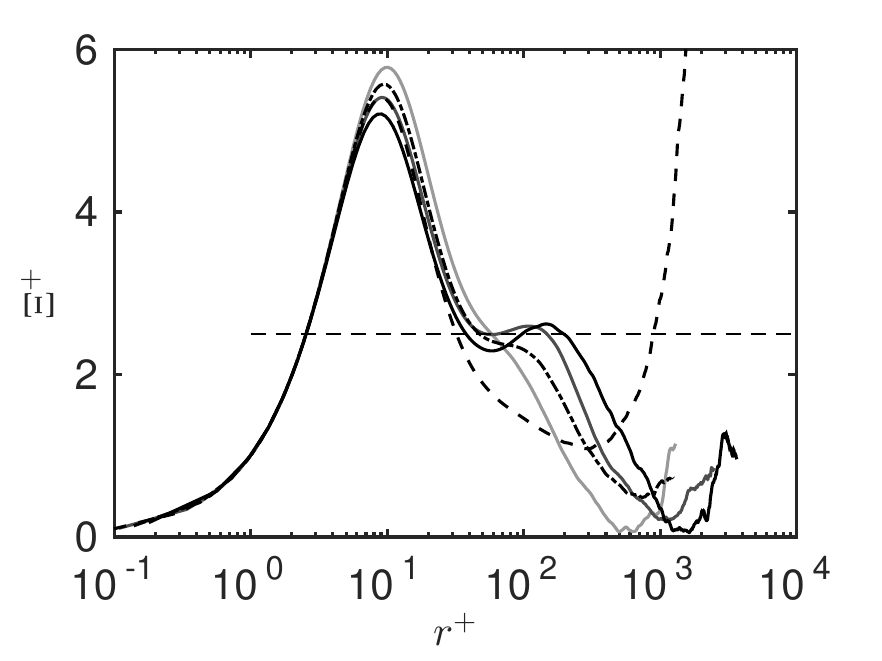}\\
  \caption{ The two top panels show the mean azimuthal (streamwise) velocity profile at the inner (left) and outer (right) cylinders in wall units. Thin dashed lines are $U^+=r^+$ and $U^+=2.5\log(r^+)+5.2$. The bottom panels show the logarithmic diagnostic function $\Xi^+$ at the inner (left) and outer (right) cylinders. The dashed straight line corresponds to $\Xi^+=2.5$ on both panels. Symbols: R1-R3 solid curves from light gray (R1) to black (R3), E5 black dashed curve, AF black dash-dot curve. }
\label{fig:q1xiwall}
\end{figure}

In this subsection we analyse the mean streamwise profiles, first with a focus on the small-gap simulations and later on the large-gap case. The top two panels of Figure \ref{fig:q1xiwall} show the mean azimuthal velocity profiles at the inner and outer cylinder in wall units for all cases. In the top two panels, a viscous sub-layer and a logarithmic layer can be seen. A classic law-of-the-wall with von Karman constants of $\kappa=0.4$ and $B=5.2$ is shown. For small gap simulations, for $r^+<200$ the match is quite good, while in the outer region, large deviations can seen. For the E5 case, the local slope in the log-layer is very different, highlighting the important effect of the curvature. Indeed, in \cite{ost14} it was shown that for $\eta=0.714$, the inverse slope in the logarithmic region was quite different from $\kappa=0.4$.

The bottom two panels of Figure \ref{fig:q1xiwall} show the logarithmic diagnostic function
\begin{equation}
\Xi^+= r^+ \displaystyle\frac{dU^+}{dr^+}.
\end{equation}

\noindent If $U^+$ is logarithmic, then $\Xi^+$ should be horizontal and equal to the inverse of $\kappa$. This is shown as a dashed line in the plots. For channels, deviations from the universal von Karman law were proposed by \cite{jim07} (based on the overlap arguments of \cite{afz73}) to have the following shape: 

\begin{equation}
 \Xi^+ = \kappa^{-1} + \alpha y/h + \beta Re_\tau^{-1/2},
 \label{eq:xichannel}
\end{equation}

\noindent where $h$ is the channel half-gap, and $\alpha=1$ and $\beta=150$ were obtained from fits to existing data. \cite{ber14} found that up to $Re_\tau=4000$ equation (\ref{eq:xichannel}) well fits the data, albeit with slightly modified constants, i.e.\ $\alpha=1.15$ and $\beta=180$. However, for PC, \cite{pir14} were not able to quantify the deviations of $\Xi^+$ in a systematic way. 

In the case of TC, fitting an equation analogous to (\ref{eq:xichannel}) for our DNS data in the logarithmic-like region does not lead to a good description of the data. The deviations in the outer region are substantial. However, a tilted S-like behaviour in $\Xi^+$ around $r^+\approx 100$ can be seen for the R2 and R3 cases, which is very similar to the one seen in the channel flow and also in PC by \cite{pir14}. 

When comparing the R1, R2 and R3 simulations, we see a strong dependence of the inner layer on $Re_\tau$. The E5 case, which we will revisit later, shows a very different structure in the near-wall region due to the large curvature. The peak in $\Xi^+$ at $r^+\approx 10$ does not saturate to a constant value across our simulations. This is probably due to the presence of the underlying large scale structures. The effect of curvature is also different from what the effective theory by \cite{gro14} predicts, presumably again due to the lack of the underlying larger scale structures in that theory, and due to the employed closure assumptions. We are unable to draw conclusions about the deviations of $\Xi^+$ based only on our DNS data. 

For the small-gap simulations, the profiles begin to deviate substantially from the logarithmic law in the bulk, i.e.\  the region with almost constant angular momentum. The deviations become especially significant from $r^+>500$ which corresponds to one tenth of the gap, or only $1\%$ curvature. The physics of TC flow in this region is dominated by the large-scale rolls.  The rolls essentially redistribute angular momentum, and this results in the quasi-flat profile for $\eta=0.909$, and the different slope in the profile for $\eta=0.5$.  The resulting streamwise mean profiles are very different from those seen in PC flow by \cite{ber14}. We note that TC flow is not Galilean invariant, and the differences between both systems could also be due to the effect of a mean rotation. However, in the case of $\eta=0.909$ this mean rotation is very small, and we may attribute the flattening of the profiles to the Taylor rolls. Indeed, the collapse, and smooth transition of the streamwise velocity profiles seen in \cite{bra15} only happens for the simulations with a large mean rotation. 

We thus may take $r^+\approx 0.1 Re_\tau$ (i.e. $\tilde{r}=0.05$) as an upper bound for the log-layer in TC for $\eta=0.909$, as for larger distances from the wall the development of the log-layer is constrained by the uniform angular momentum resulting from the presence of the rolls. This means that the effective $Re_\tau$ of the simulations decreases substantially, and the possible logarithmic regions extend less into the bulk. If one takes a lower bound for the start of the logarithmic layer to be approximately at $r^+=3Re_{\tau}^{1/2}$ \citep{mar13} and an upper bound at $r^+\approx 0.1 Re_\tau$ due to the dominance of curvature effects and of the rolls, we obtain that the logarithmic layer extends in the $r^+$ range of $150<r^+<400$. This is a small range, which  does not provide a sufficient separation of scales to accomodate for the near-wall cascades which result in a well-developed logarithmic profile. We can take the less conservative lower bound for the start of the log-layer of $r^+\approx 30$ \cite{pop00}, but even with this bound, the range of length scales is still insufficient. 

\begin{figure}
  \centering
  \includegraphics[width=0.49\textwidth]{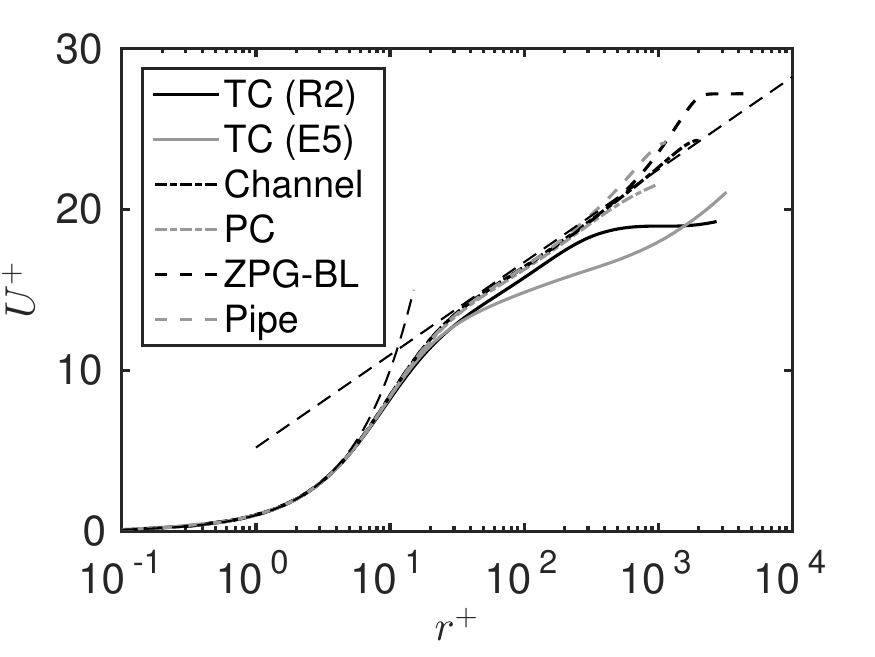}
  \includegraphics[width=0.49\textwidth]{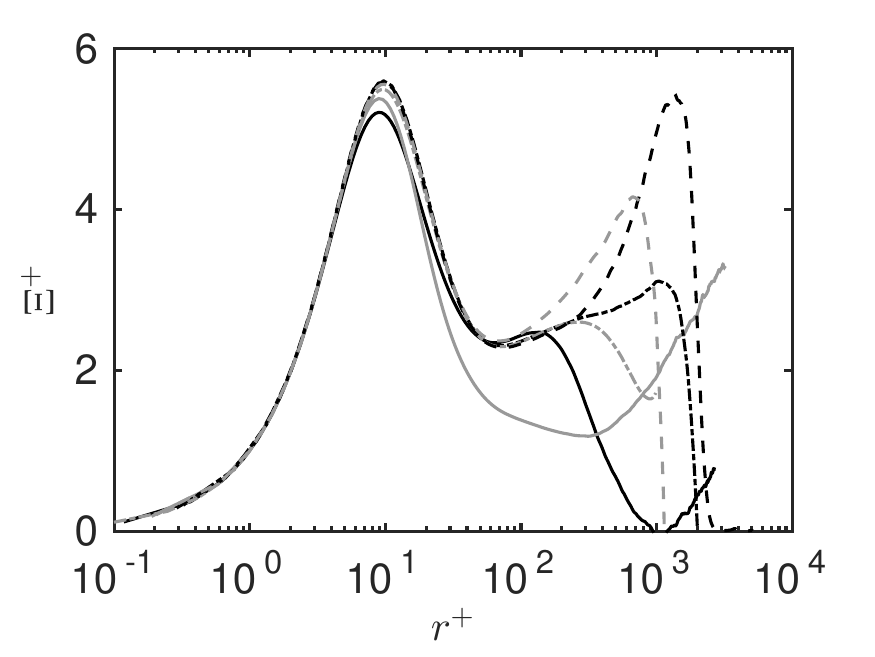}\\
  \caption{ Left panel: mean streamwise component in wall units for several canonical flows. Right panel: $\Xi^+$ for the same flows.}
\label{fig:meancomp}
\end{figure}

Therefore, a higher $Re_\tau$ is needed in small-gap TC to see a comparable law-of-the-wall to those of channels or pipes. This is shown in more detail in Figure \ref{fig:meancomp}, which compares the streamwise mean profile and $\Xi^+$ in several canonical flows at a frictional Reynolds number of $Re_\tau\approx1000$. TC shows a smaller peak value of $\Xi^+$ when compared to the other canonical flows. However, the near-wall region of small-gap TC is still surprisingly similar to other canonical flows, given how different all flows are. On the other hand, large-gap TC shows significant deviations, something we can expect as the physics is dominated by the curvature and the centrifugal instabilities. Finally, we note that we cannot conclude that we observe a logarithmic layer in our simulations, even if we may speculate based on experimental results \citep{hui13} that further DNS at larger $Re_\tau$ will lead to the development of a near wall region with comparable properties to those of other canonical flows. 

As mentioned previously, the large-gap case shows a very different behaviour (cf. all panels of figure \ref{fig:q1xiwall}). Other approaches to model this behaviour can be taken. \cite{gro14} attempt to calculate the near-wall streamwise profile in Taylor-Couette flow following the spirit of Prandtl's law-of-the-wall. They use a turbulent diffusivity to account for the extra mixing provided by turbulence. The resulting equations predict a logarithmic-layer for the \emph{angular} velocity with a curvature correction factor $A$ equal to $A=(1+(r-r_i)/r_i)^{-3}$ for the inner cylinder and $A=(1+(r_o-r)/r_o)^{-3}$ for the outer cylinder. Again we stress that this approach does not consider the effect of the Taylor rolls, which efficiently mix angular momentum in the bulk, so it can only be valid in the near-wall region. 

\begin{figure}
  \centering
  \includegraphics[width=0.49\textwidth]{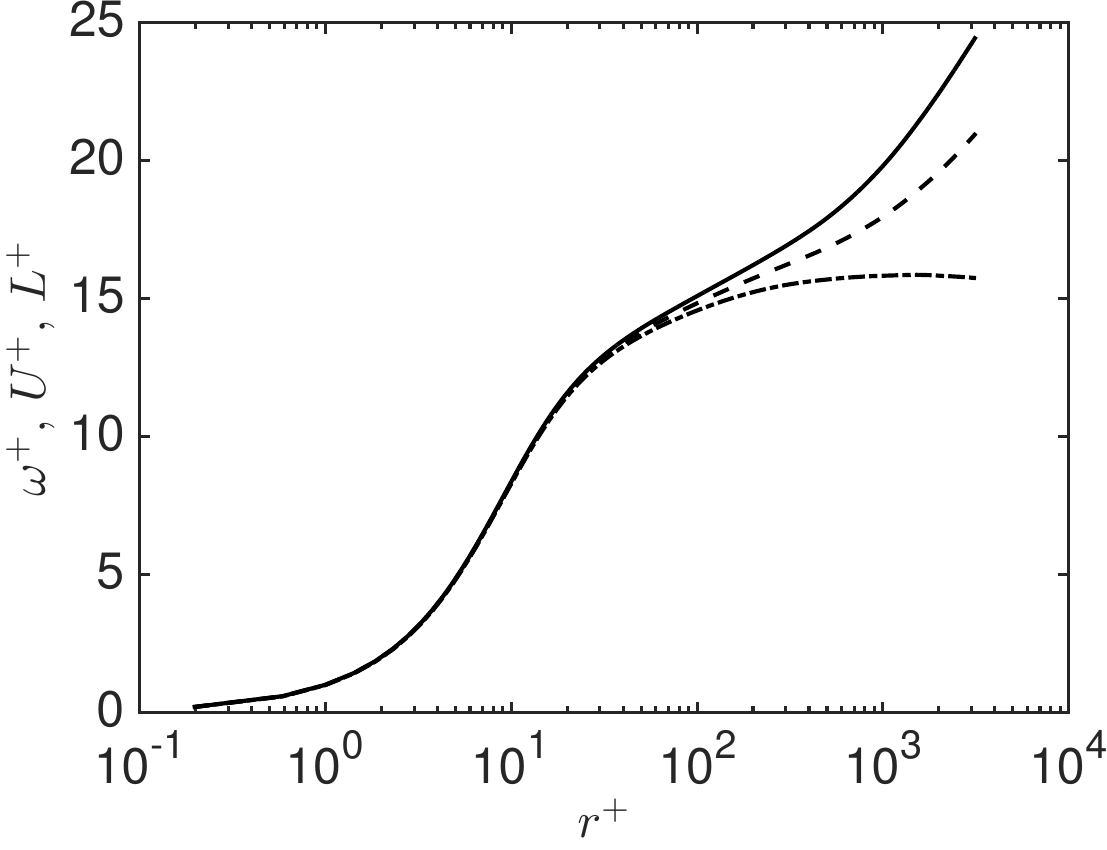}
  \includegraphics[width=0.49\textwidth]{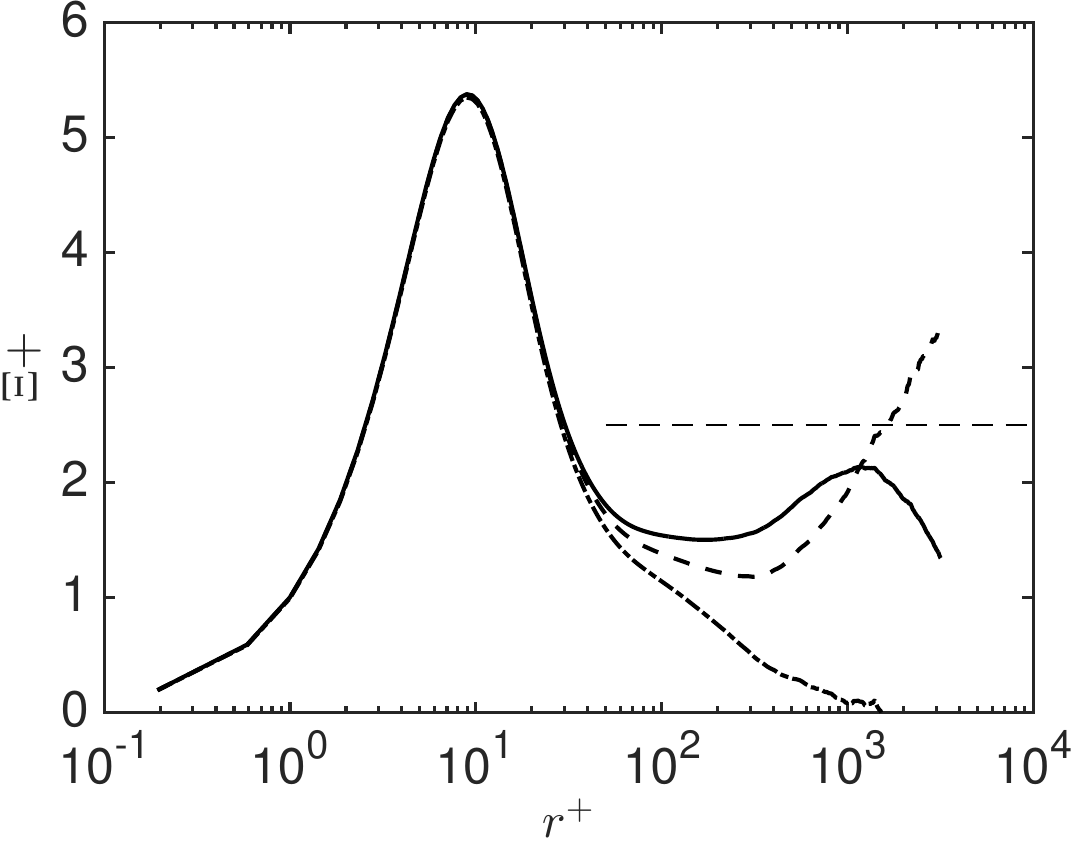}\\
  \caption{ Left: Comparison between angular momentum (dash-dot), azimuthal velocity (dashed) and angular velocity with curvature corrections (solid curve) in wall-units for the E5 case. Right: Local logarithmic slope for the variables in the left panel. The horizontal line corresponds to an inverse logarithmic slope of $\kappa=0.4$.  }
\label{fig:eta05}
\end{figure}

In the left panel of figure \ref{fig:eta05}, we show the profiles of angular velocity (with curvature corrections), azimuthal velocity and angular momentum in wall units at the inner cylinder for the E5 case, in order to compare the consistency of the profiles with a logarithmic layer. To further illustrate this, the right panel shows the local logarithmic slope (or logarithmic diagnostic function) of the three variables. In the near-wall region, the profiles are indistinguishable in the viscous sublayer. Due to the minuscule size of $\delta_\nu$, the difference between angular momentum and angular velocity is less than $1\%$ at $r^+=20$ ($\tilde{r}\approx0.003$) The profiles only noticeably deviate from each other for $r^+>50$. Indeed, in this region, the angular velocity shows the flattest diagnostic function, consistent with the predictions of \cite{gro14}, and the numerical simulations of \cite{ost14} at $\eta=0.714$. However, this log-layer does not extend further than $r^+=300$, i.e.\ one twentieth of the gap width. Here, the profiles become consistent with the constant angular momentum bulk $\bar{L}=0.5$, and we see the diagnostic function for $L^+$ becomes zero. Therefore, we conclude that even though logarithmic profiles appear for the angular velocity, again these are limited in size to little more than half a decade in $r^+$, and do not possess sufficient range of length scales to sustain the near-wall cascades. We also note that for the E5 case, the mean rotation of the system is quite strong, and this may also play an effect on the velocity profile, causing the discrepancies in the bulk region.

\subsection{Velocity and pressure fluctuations}
\label{sec:fluct}

\begin{figure}
  \centering
  \includegraphics[width=0.49\textwidth]{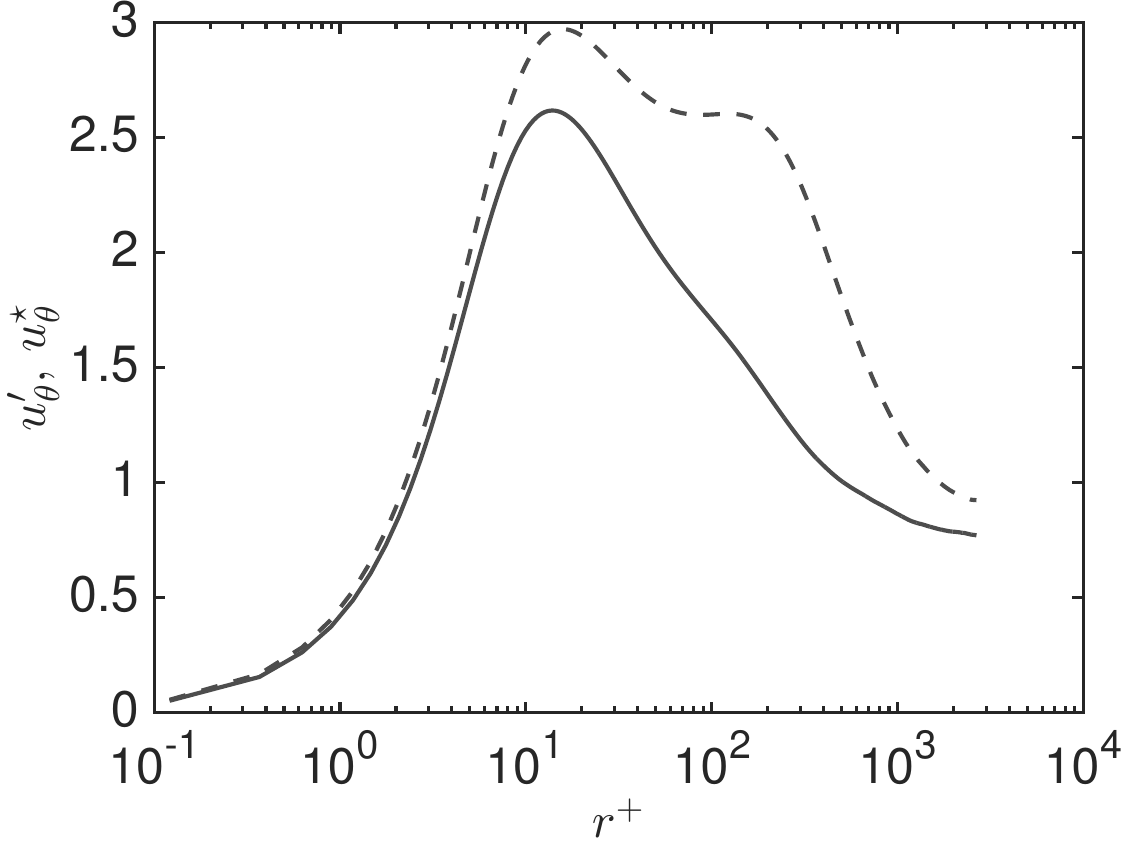}
  \includegraphics[width=0.49\textwidth]{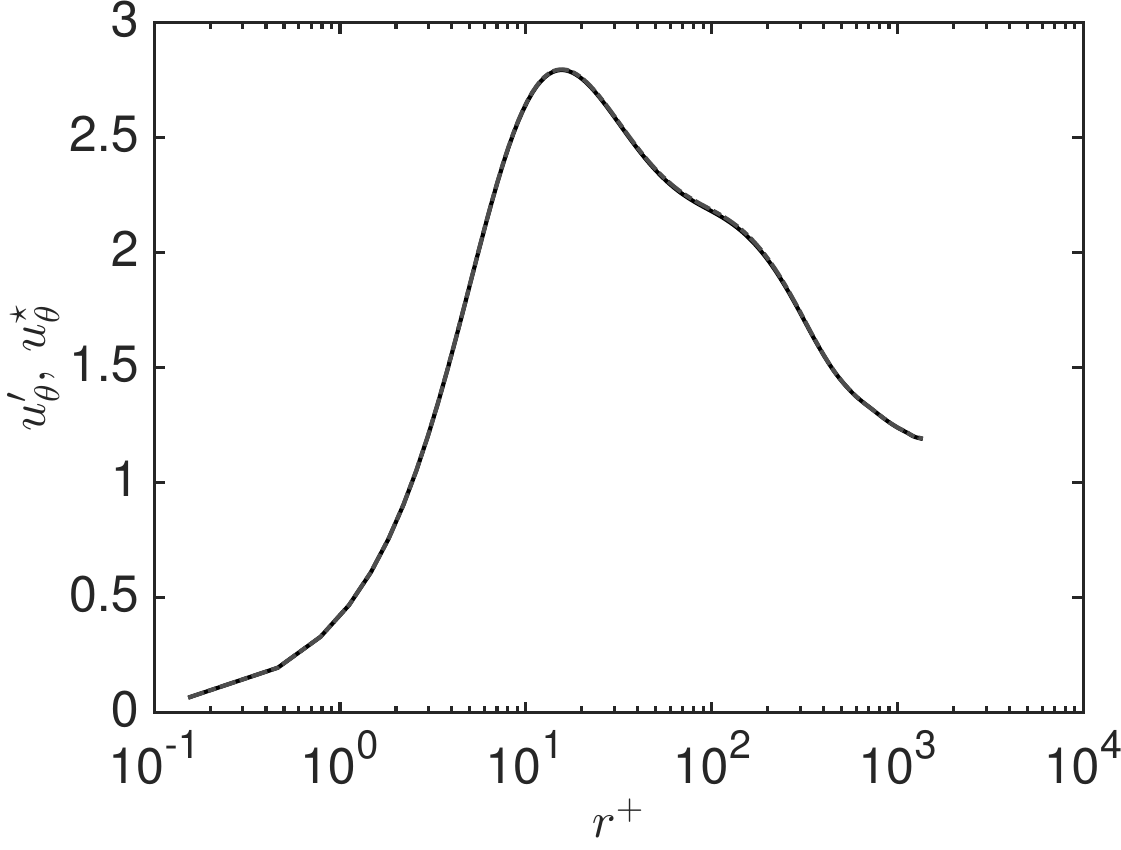}
  \caption{ Mean fluctuation profiles for azimuthal velocity for R2 case with axial inhomogeneity (left) and AF case with no fixed axial inhomogeneity (right). Symbols: $u_\theta^\prime$ dashed curve, $u^\star_\theta$ solid curve.}
\label{fig:twoways}
\end{figure}

As mentioned in Section \ref{sec:numset}, we have defined two ways to quantify the fluctuations. Figure \ref{fig:twoways} shows the resulting difference between both ways for both the R2 case (with fixed rolls and axial inhomogeneity) and the AF case (with no fixed rolls, and thus no axial inhomogeneity). While for the R2 case, very significant differences can be seen- including the existence (or not) of a second hump, for the AF case the curves collapse on one another and the difference is beneath the temporal convergence error. 

To understand this, we must first distinguish the two types of temporal and spatial dependence of the velocity (and pressure) fields seen in small-gap TC: the fast small-scale fluctuations inside the rolls, due to the herringbone-like streaks, and the large-scale, quasi-stationary axial dependence due to the presence of the Taylor rolls. Provided that the rolls are static in the axial direction (cases R1-R3) due to the order of the averaging and subtraction operations only the small-scale fast fluctuations affect $\phi^\star$. The rolls segregate streak detachment regions and streak impacting regions, which have very different velocities, and this spatial dependence is absent using this definition. This is not the case if the rolls are convected (AF simulation) or disappear, (E5 simulation). Here, the axial dependence of the flow disappears when computing the mean field, and then both small- and large-scale fluctuations are computed when calculating both $\phi^\prime$ and $\phi^\star$, and so they coincide.

\begin{figure}
  \centering
  \includegraphics[width=0.49\textwidth]{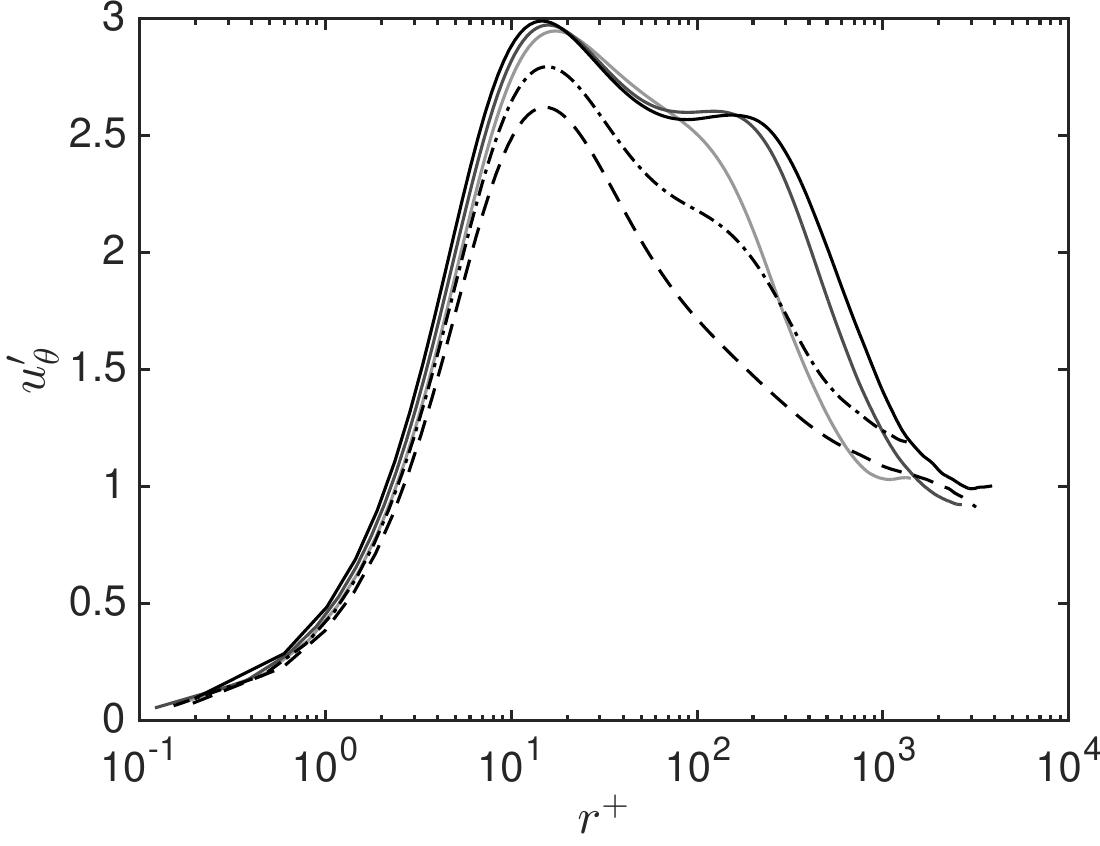}
    \includegraphics[width=0.49\textwidth]{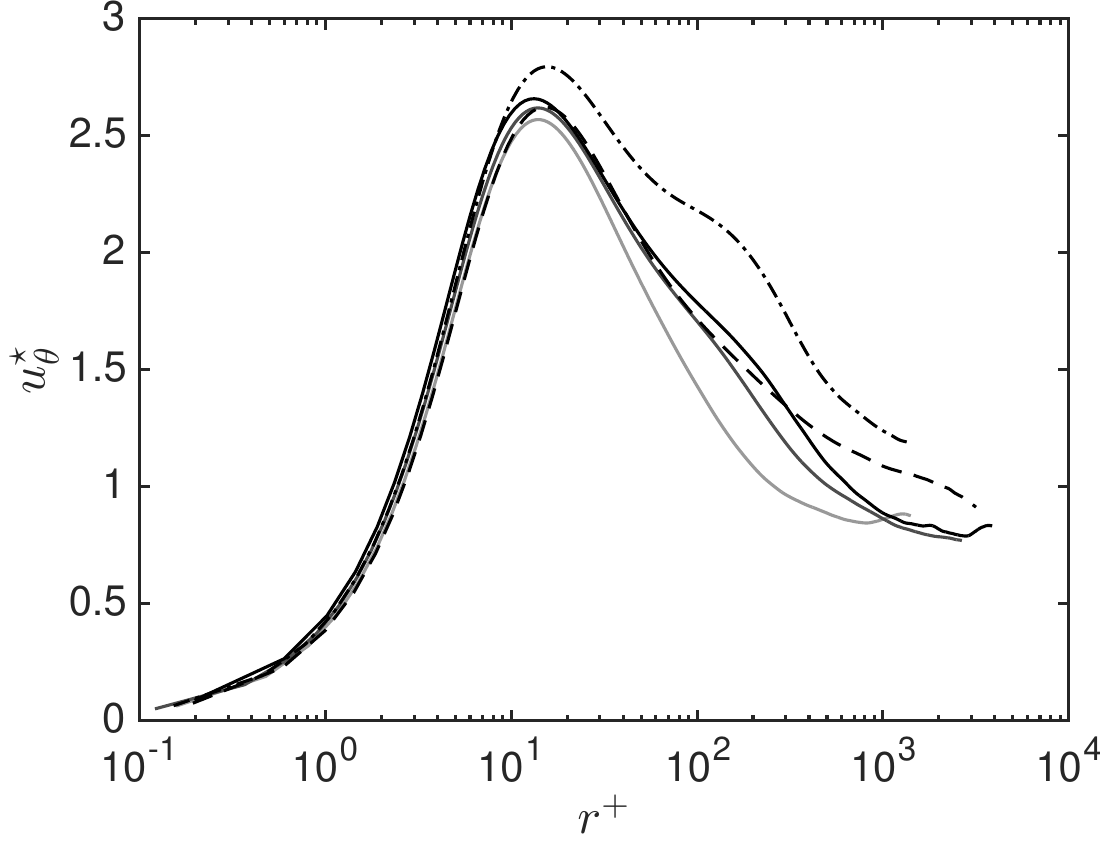}\\
  \includegraphics[width=0.49\textwidth]{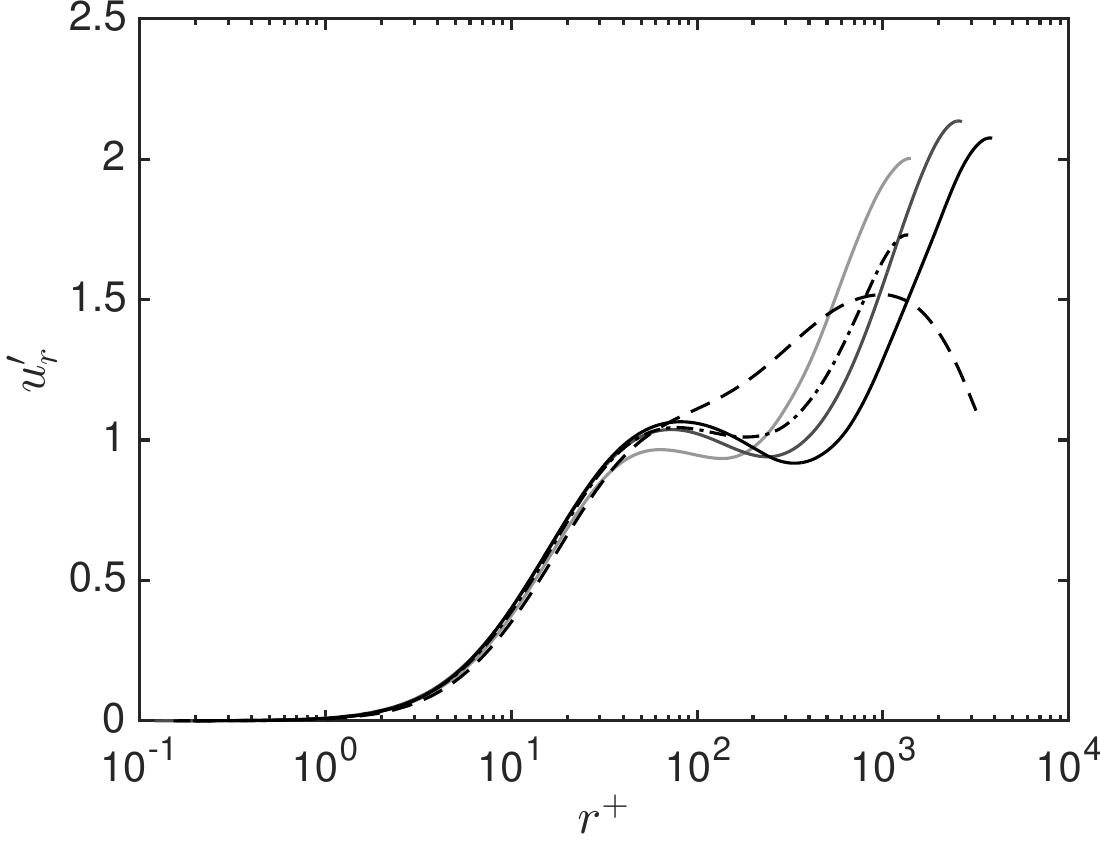}
    \includegraphics[width=0.49\textwidth]{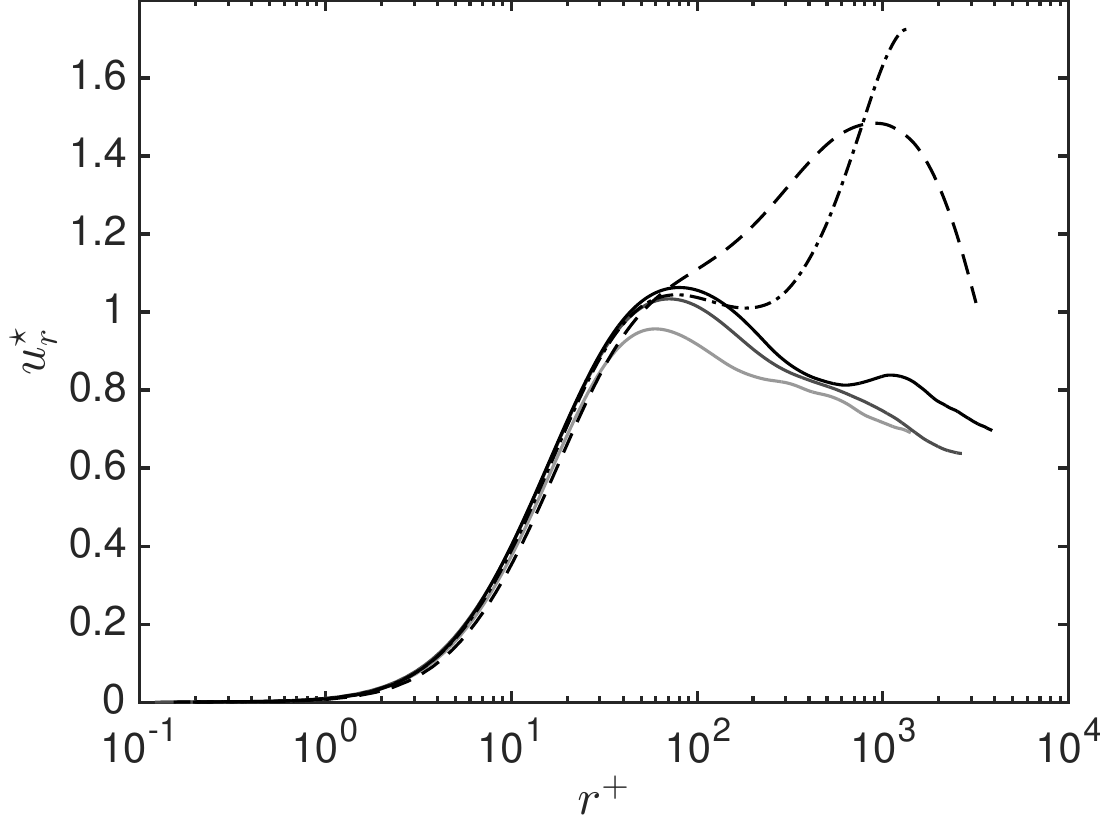}\\
  \includegraphics[width=0.49\textwidth]{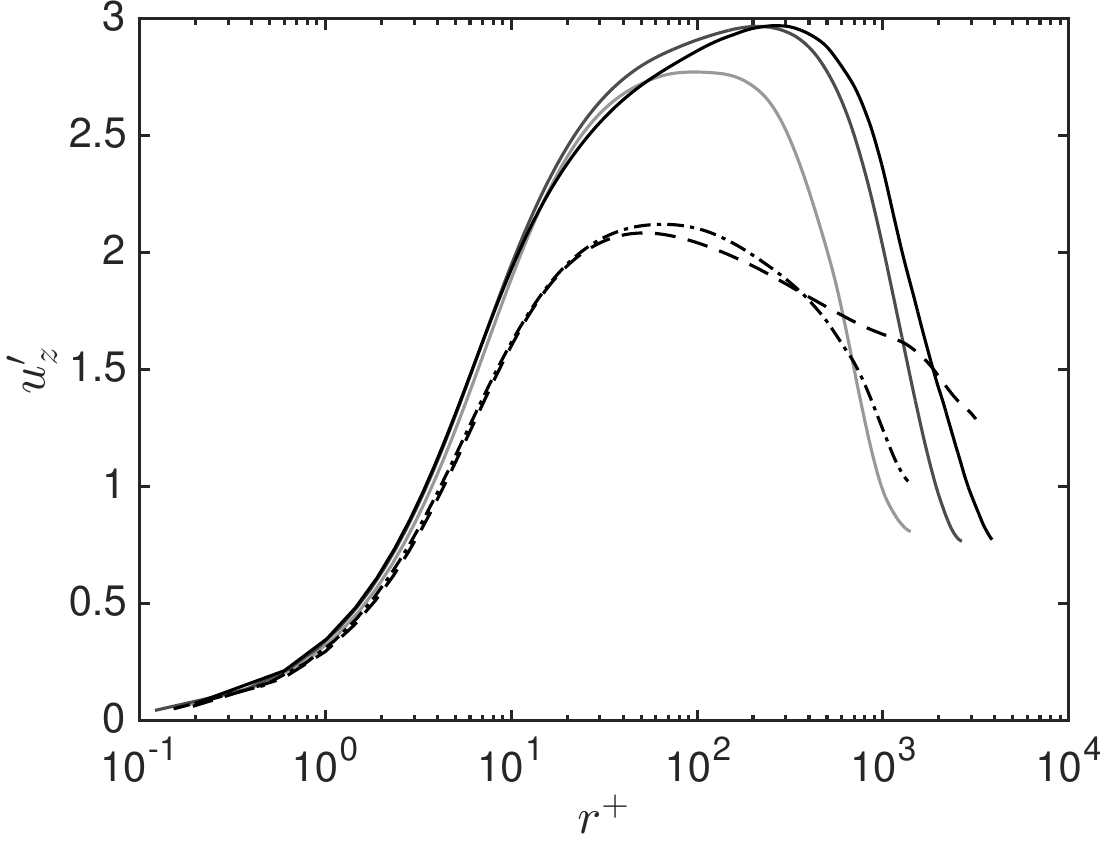}
    \includegraphics[width=0.49\textwidth]{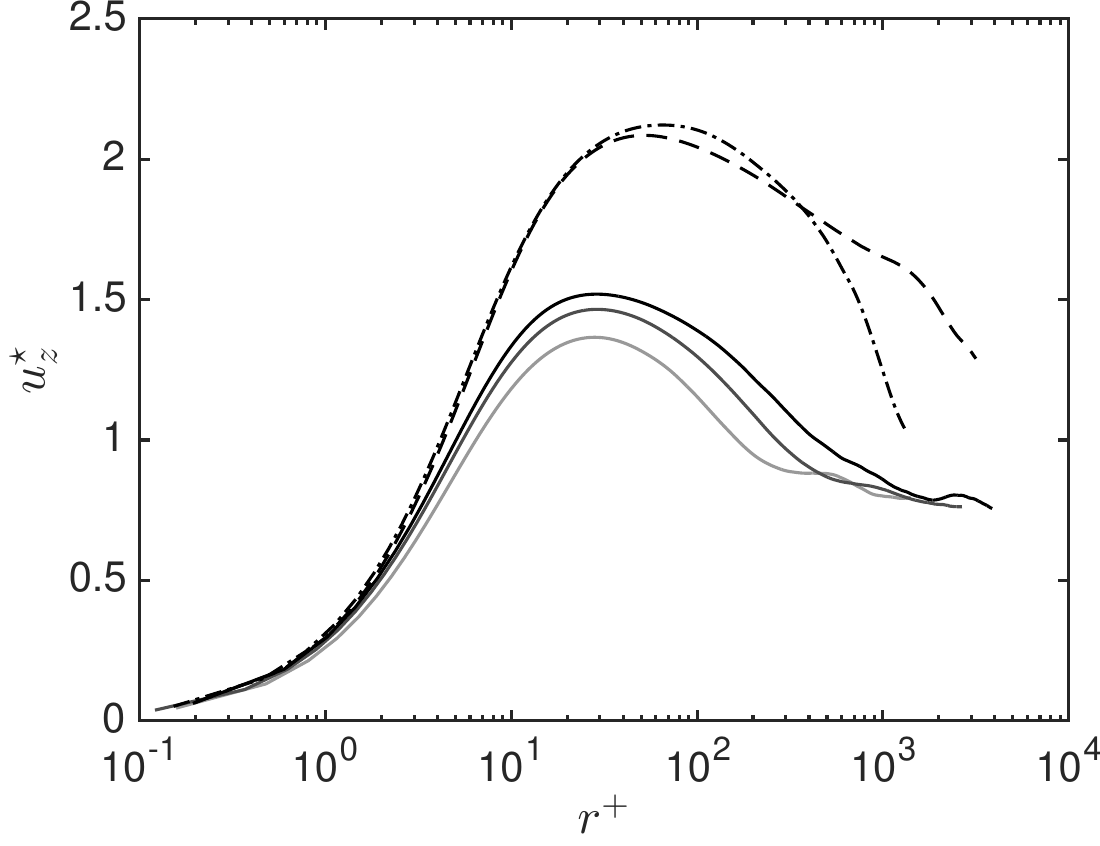}\\
  \caption{ Fluctuation profiles using both types of definition for velocities and pressure at the inner cylinder in wall units.  Symbols: R1-R3 solid curves from light gray (R1) to black (R3), E5 black dashed curve, AF black dash-dot curve. }
\label{fig:primewall}
\end{figure}

\begin{figure}
  \centering
  \includegraphics[width=0.65\textwidth]{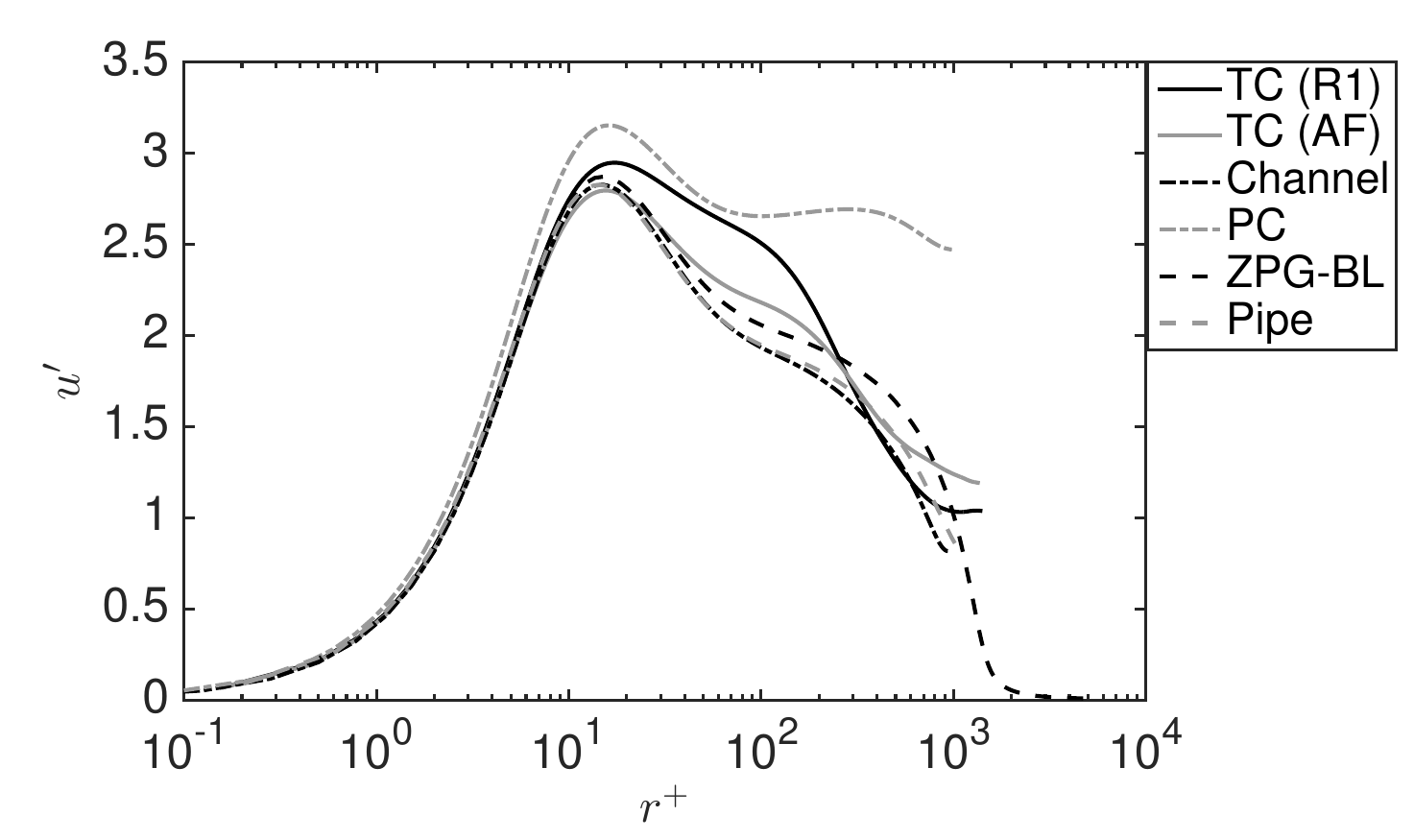}
  \caption{ Streamwise velocity (azimuthal velocity for TC) fluctuations in wall units for the R1 and AF cases at the inner cylinder, and some canonical flows. }
\label{fig:plusfig2}
\end{figure}

Figures \ref{fig:primewall} shows the velocity fluctuations at the inner cylinder. For the R1-R3 cases, the velocity fluctuations are generally either higher (when computed using $\phi^\prime$) or lower (when computed using $\phi^\star$) than those seen for channels at comparable $Re_\tau$ (cf. \cite{pir14} or Figure \ref{fig:plusfig2}). Except for the AF case, there is no presence of a second hump in $\phi^\star$, and the level of fluctuations slightly increases with $Re_\tau$, as expected. The difference between the two ways of quantifying the fluctuations is apparent here. The axial inhomogeneity causes a significant increase of the measurement, especially in the bulk region, where the structure is more pronounced.

\begin{figure}
  \centering
  \includegraphics[width=0.49\textwidth]{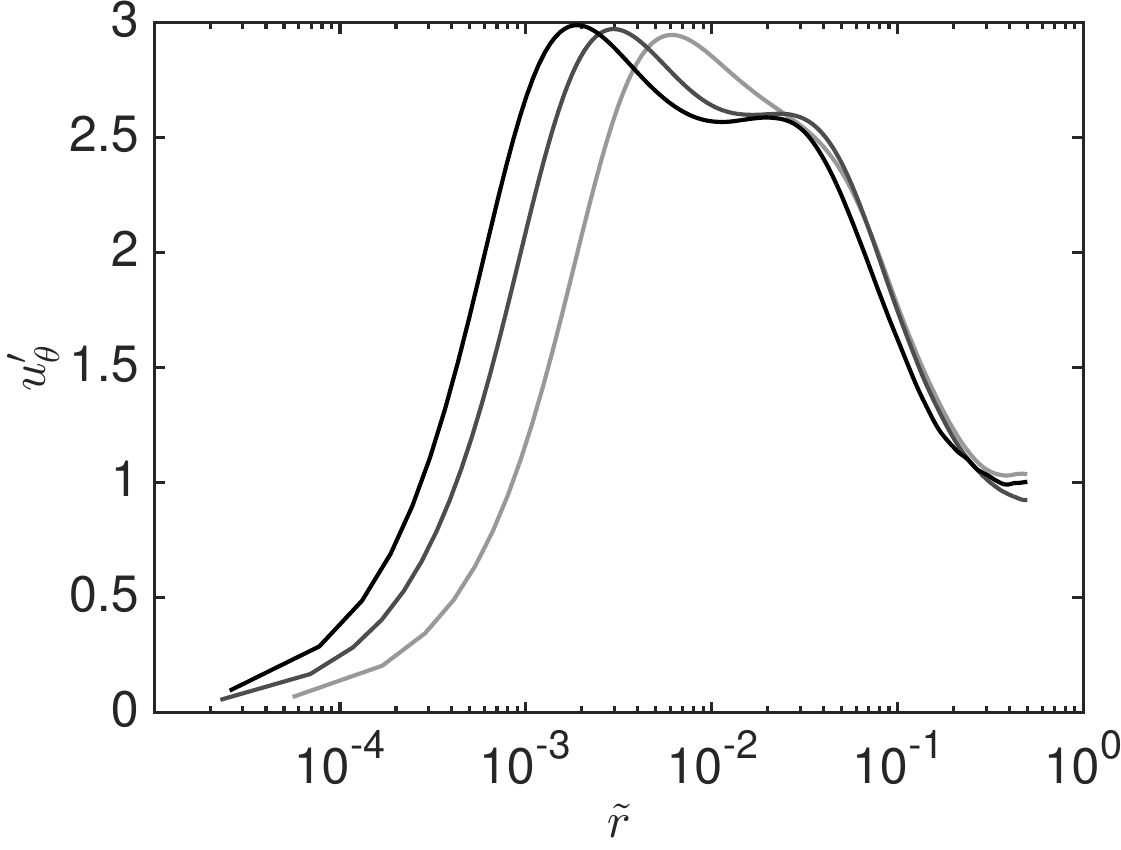}
  \includegraphics[width=0.49\textwidth]{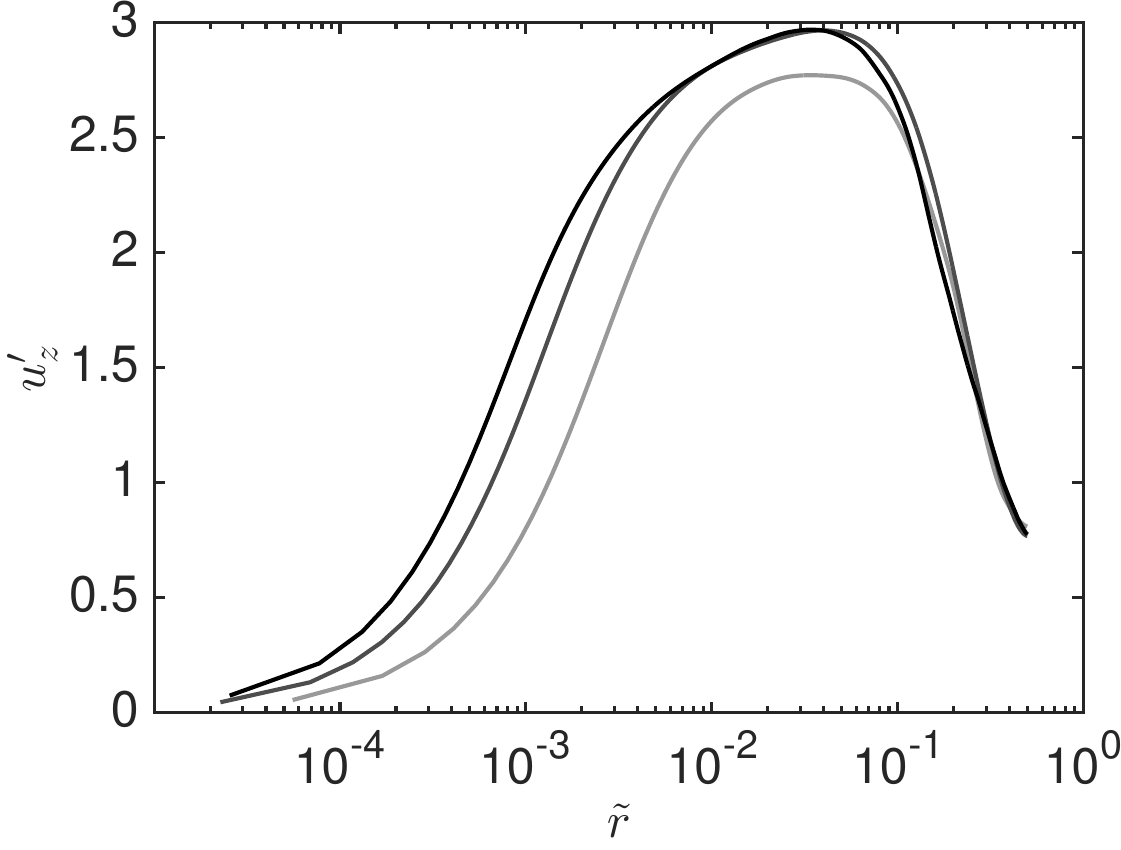}\\
    \includegraphics[width=0.49\textwidth]{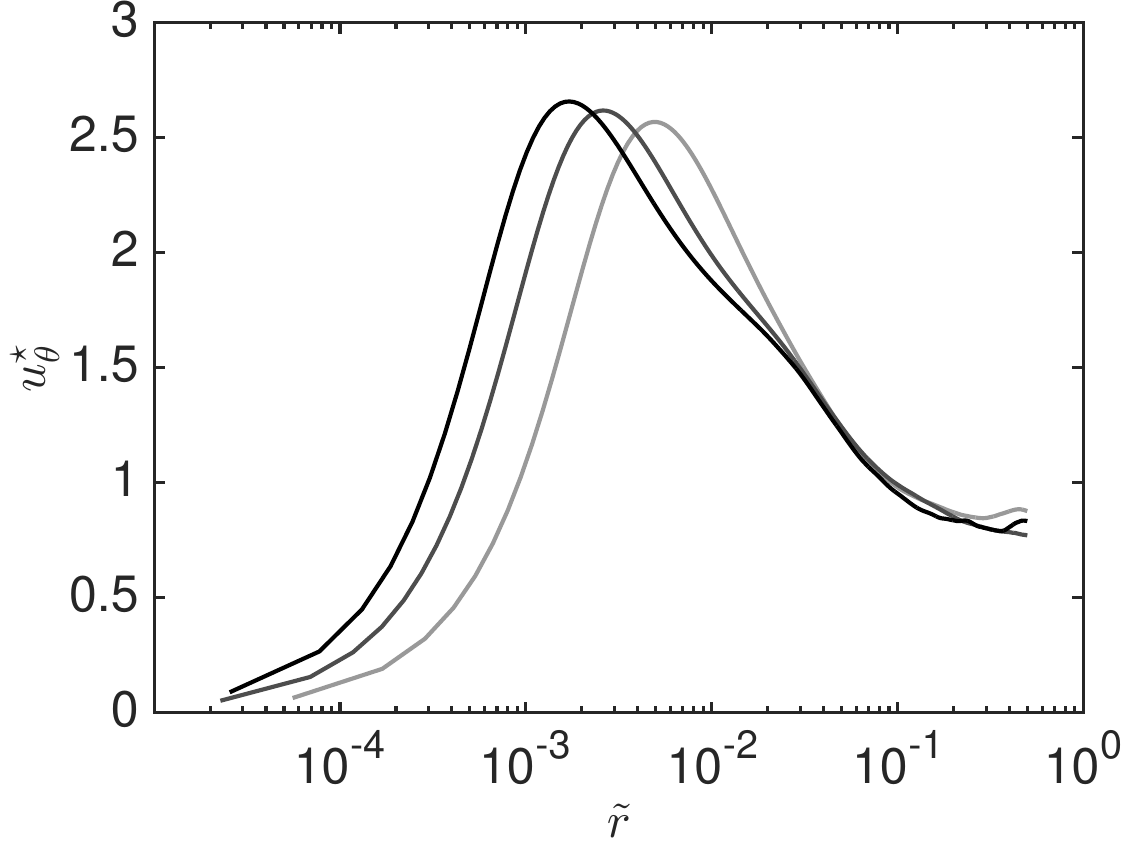}
      \includegraphics[width=0.49\textwidth]{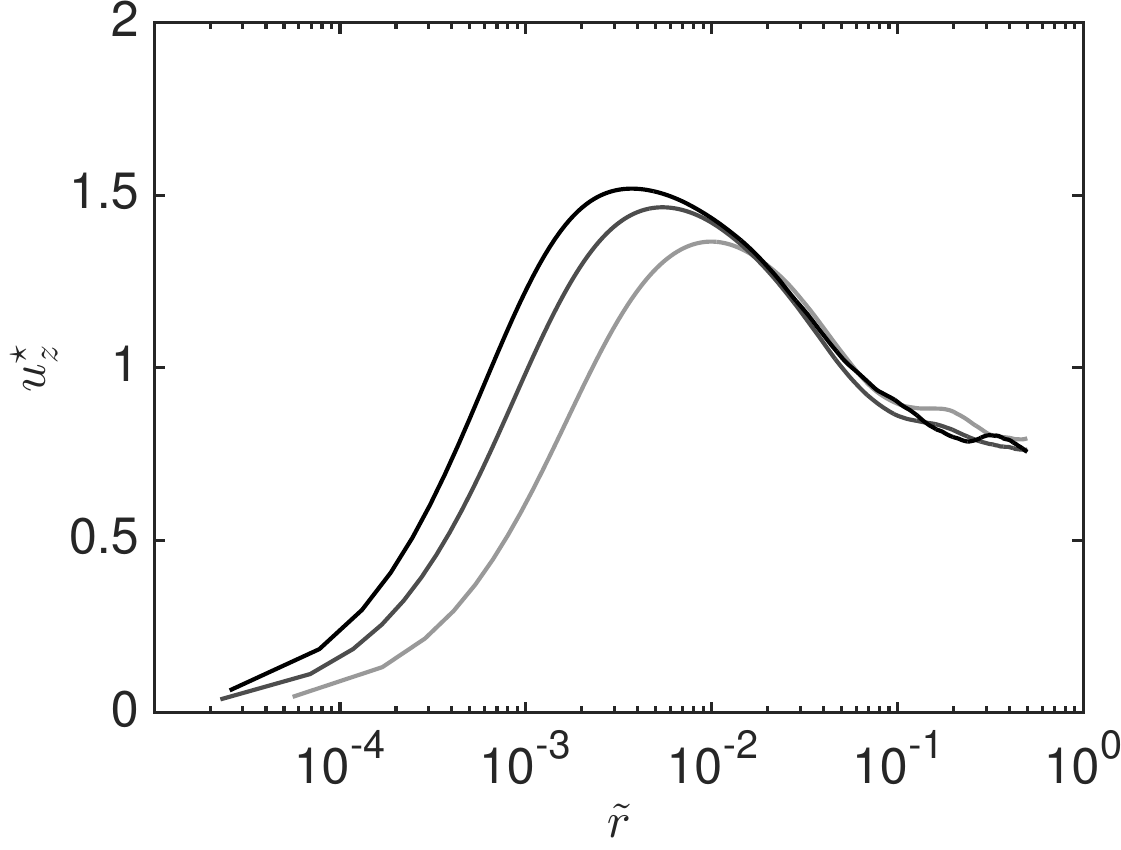}
  \caption{ Fluctuations of the azimuthal (left panels) and axial (right panels) velocity components for the R1, R2 and R3 cases in outer units. Regardless of the way of calculating the fluctuations, an overlap region, where the fluctuations collapse can be seen in both panels. Symbols: R1-R3 solid curves from light gray (R1) to black (R3).  }
\label{fig:plusouter}
\end{figure}

Velocity fluctuations are also shown with the horizontal axis in outer units in Figure \ref{fig:plusouter}. A reasonable collapse can be seen in an ``overlap'' layer. This layer is defined between $r^+>100$ and $\tilde{r}<0.2$, i.e. $r^+<0.4 Re_\tau$. Here, $u^\prime_\theta$ and $u^\prime_z$ show a local profile which is consistent with logarithmic behaviour. This can mean that a region in which overlap arguments are valid exists in TC. However, DNS with larger $Re_\tau$ is needed to sufficiently decouple the scales, and obtain convincing evidence for this overlap layer, with logarithmic profiles both in the mean velocities and the pressure and velocity fluctuations.  

This is further quantified in figure \ref{fig:plusfig2} which compares the azimuthal (streamwise) velocity fluctuations at the inner cylinder for the R1 and AF cases with those of the streamwise velocity for several canonical flows at around $Re_\tau\approx 1000$. Except for PC and the R1 case, a remarkable agreement in the value of $u^\prime$ at the peak of $r^+\approx 12$ is obtained, including the data from the AF case. The R1 case has a higher amount of fluctuations than the rest of the cases. This might be expected, both due to the reasons mentioned previously about the axial inhomogeneity (notable when comparing the AF and R1 cases), and because in TC we are driving a secondary flow due to a centrifugal instability, and this also can affect the level of fluctuations (the level is lower when comparing only the fluctuations coming from the streaks). This second effect is probably the cause for the smaller level of fluctuations seen in the E5 case even when the large scale structure is absent. The addition of an axial flow or of a large curvature increases the level of fluctuations, especially those of the radial and axial velocities. This is the reason we see both a second peak in $u^\prime_\theta$ for high $r^+$ in the AF case, and higher values of $u^\prime_r$ and $u^\prime_z$, especially near the mid-gap for both the E5 and the AF cases. 

\begin{figure}
  \includegraphics[width=0.49\textwidth]{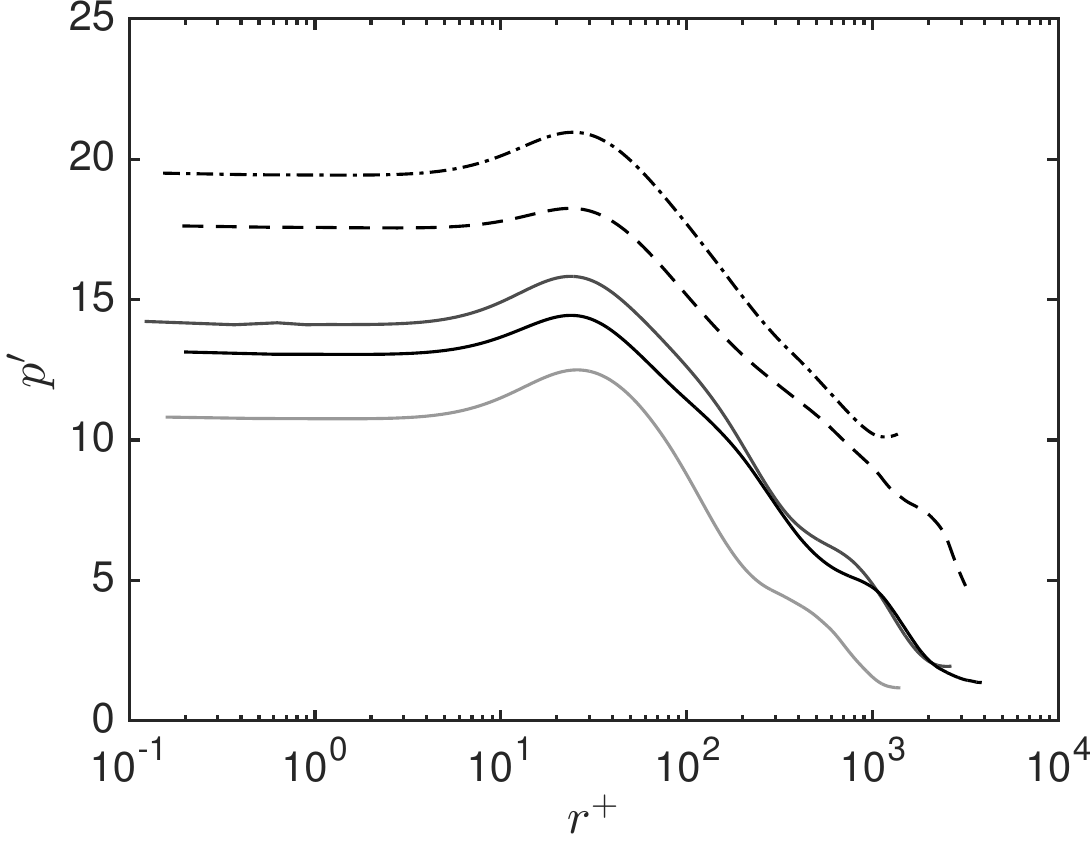}
    \includegraphics[width=0.49\textwidth]{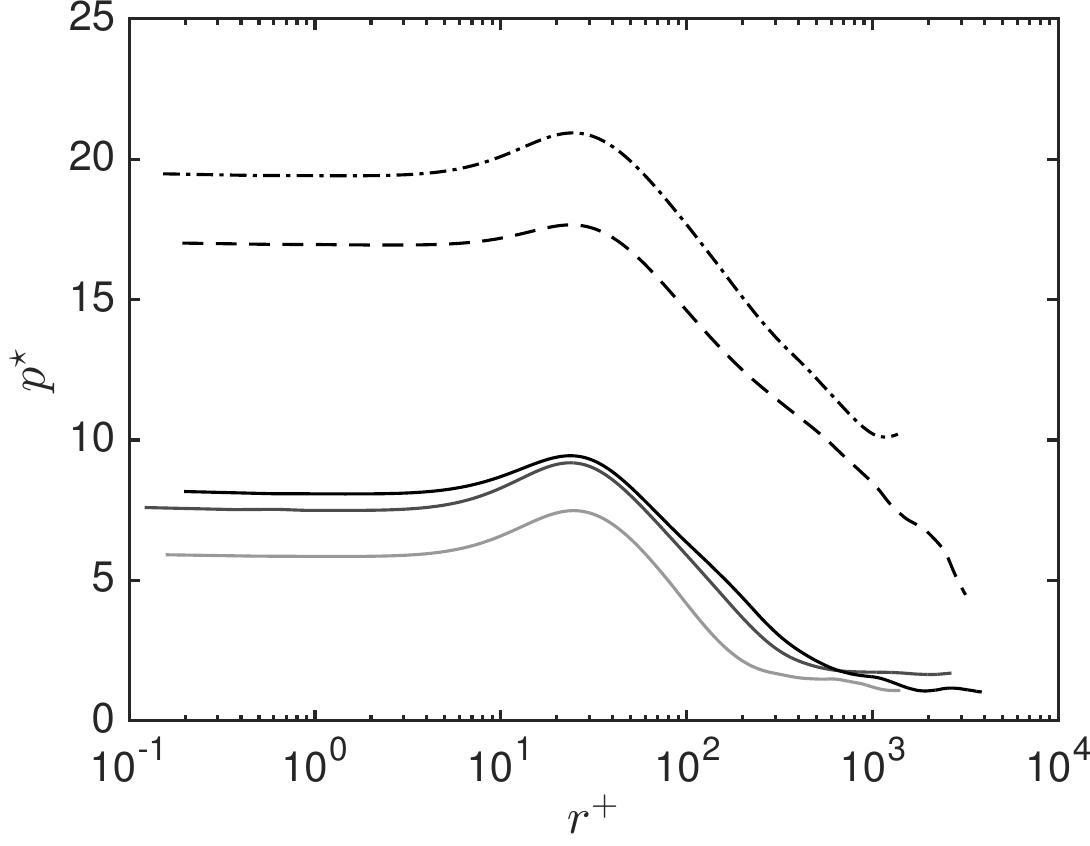}\\
      \caption{ Fluctuation profiles using both types of definition for velocities and pressure at the inner cylinder in wall units.  Symbols: R1-R3 solid curves from light gray (R1) to black (R3), E5 black dashed curve, AF black dash-dot curve. }
\label{fig:ppluswall}
\end{figure}

Clear logarithmic behaviour is also seen for $p^\prime$ in figure \ref{fig:ppluswall}, as well as a reasonable collapse of the data in this region for the R2 and R3 cases when looking at the $p^\star$ quantification. This behaviour is consistent with the existence of an overlap region. In addition, the pressure fluctuations in the AF case are almost three times larger than for the R1 case, but this is due to the way the axial pressure gradient is imposed- with a time varying magnitude.

\subsection{Velocity spectra}
\label{sub:spec}

\begin{figure}
  \centering
  \includegraphics[width=0.49\textwidth]{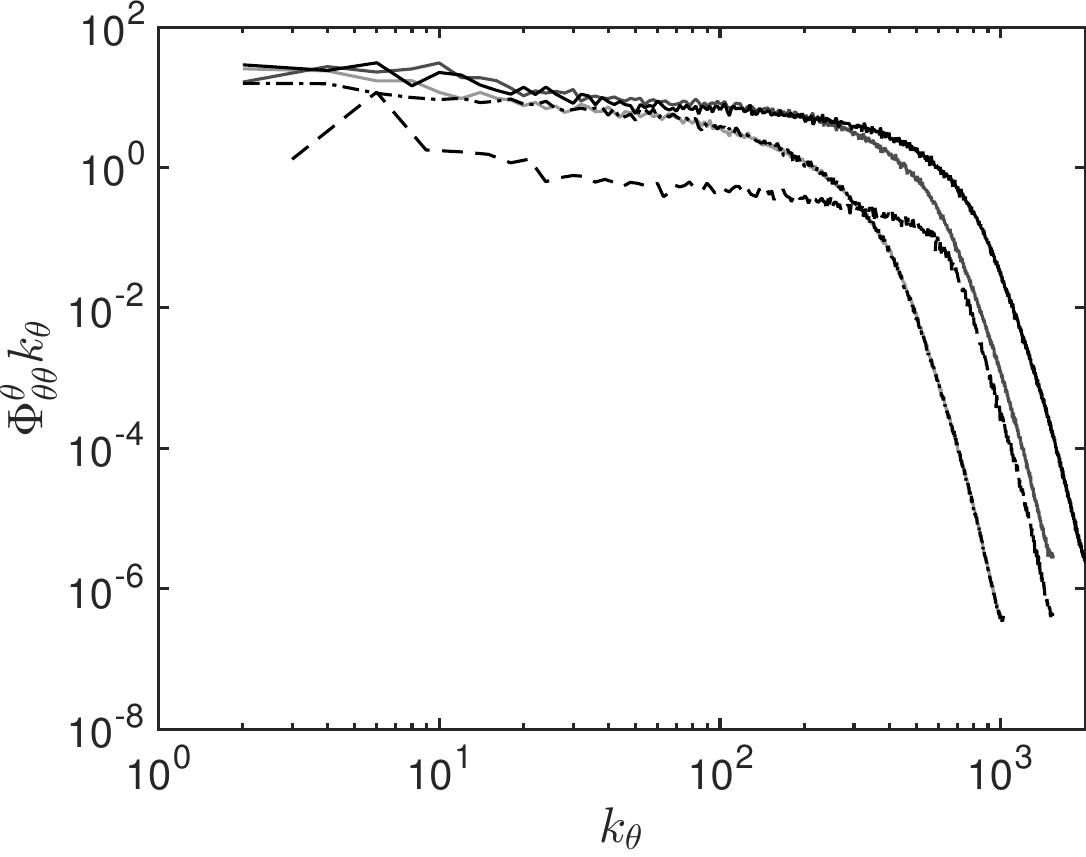}
  \includegraphics[width=0.49\textwidth]{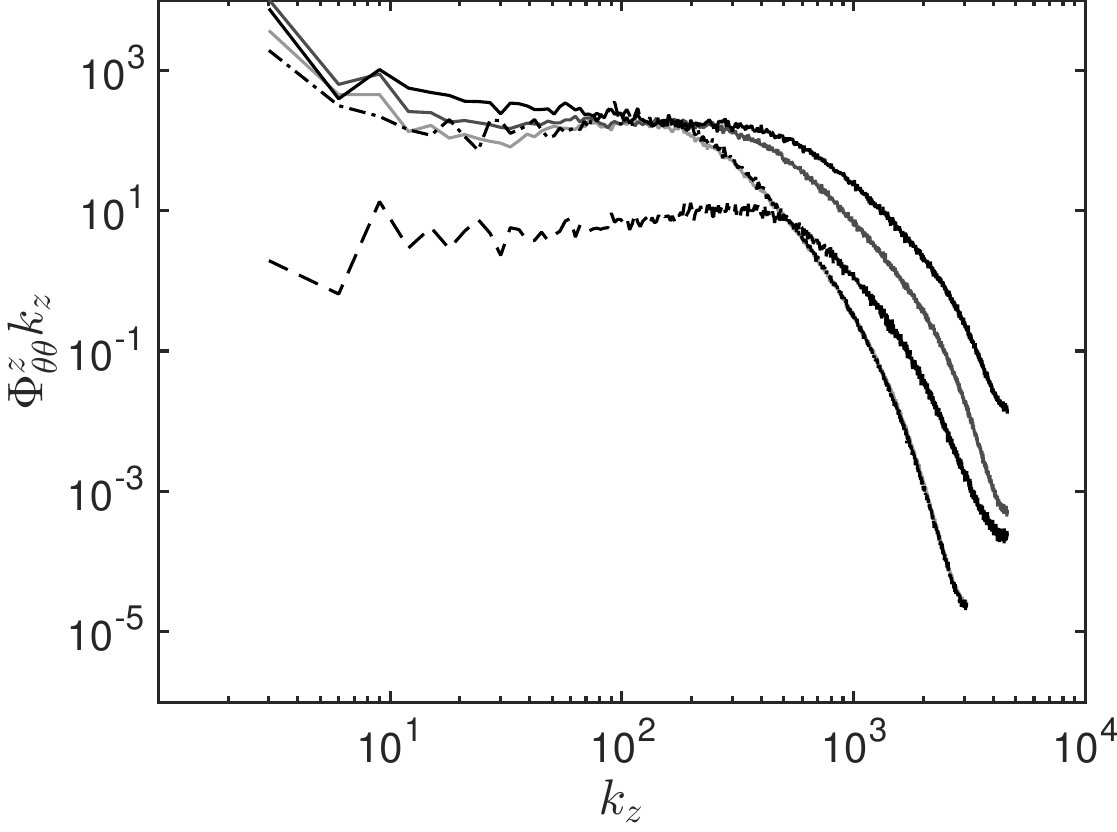}\\
  \includegraphics[width=0.49\textwidth]{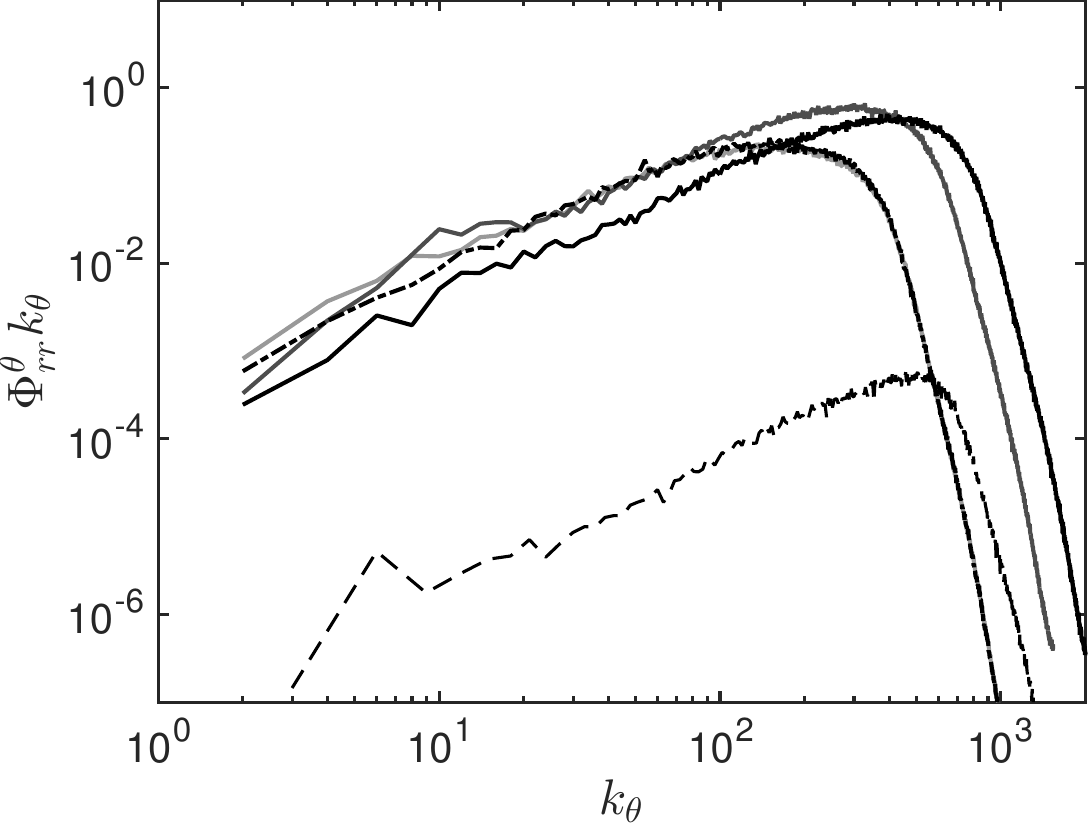}
  \includegraphics[width=0.49\textwidth]{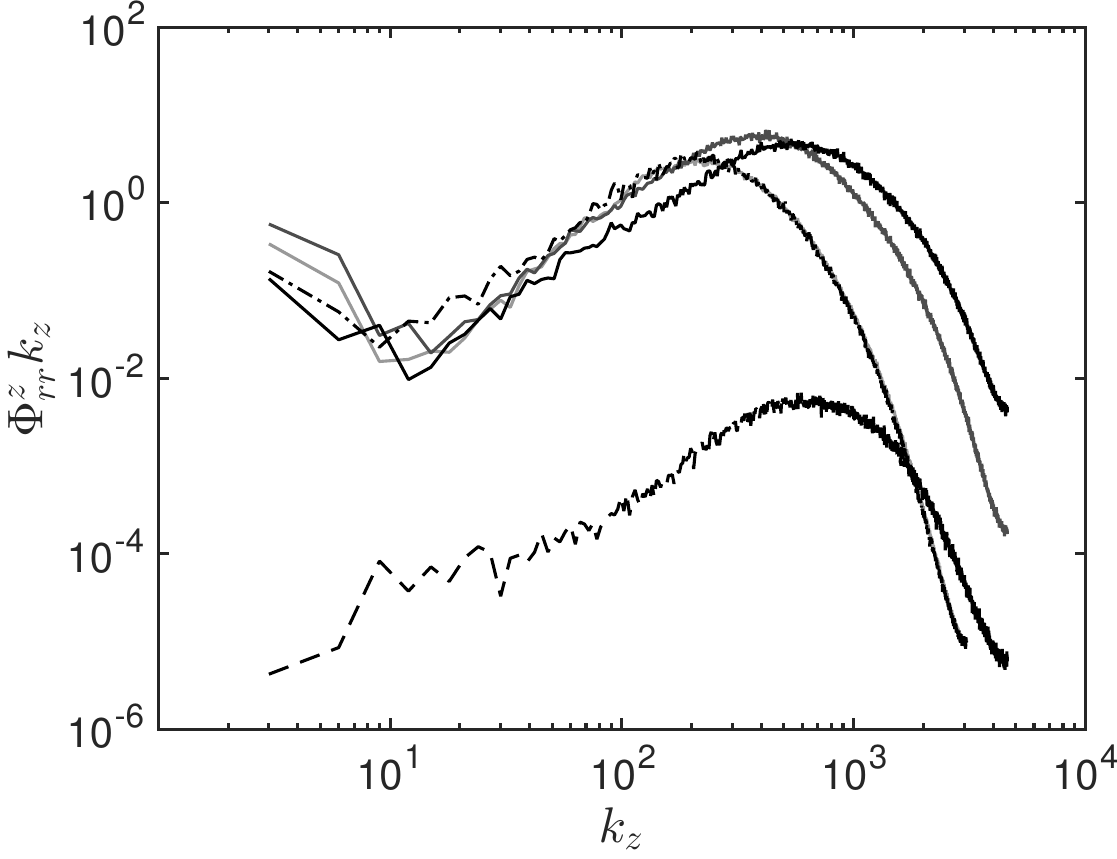}\\
  \caption{ Azimuthal (left) and axial (right) spectra for azimuthal and radial velocity components for all simulations at $r^+\approx 12$. Symbols: R1-R3 solid curves from light gray (R1) to black (R3), E5 black dashed curve, AF black dash-dot curve. }
\label{fig:slab2spectra}
\end{figure}

Taylor rolls contain a significant amount of the kinetic energy in the flow, both in the bulk and in the boundary layer. This is quantified in figure \ref{fig:slab2spectra}, which shows the azimuthal and axial spectra of the azimuthal and radial velocity for $r^+\approx 12$ (near the wall)in outer units. The Taylor rolls are associated to the lowest wavelengths in the axial spectra, and these show a prominent maximum at the fundamental wavelength. This signature is not only present for $\Phi_{\theta\theta}$ but also for the radial velocity spectra $\Phi_{rr}$ for the small gap cases. For the E5 case, this signature is not present, consistent with the fact that the rolls have faded. 

Furthermore, all small gap cases show a maximum in the cospectra $\Phi_{\theta r}$ (not plotted here) corresponding to the wavelength of axisymmetric Taylor rolls, i.e. $k_\theta=0$, $k_z=2\pi/\lambda_{TR}$. If present, the rolls dominate the convective transport of angular velocity through the Reynolds stresses in the boundary layer. This means that Taylor rolls are not inactive in the sense of \cite{tow76} and \cite{hoy06}. Rolls actively transport (or redistribute) the angular velocity current. This result, well known at low Reynolds numbers, is extended here to the high Reynolds numbers analysed. Furthermore, unlike the large scale structures in channel flow and PC flow, the rolls may be considered to be ``attached'' to the wall, and their presence is felt inside the boundary layer. For the E5 case, no signature of this large-scale roll can be seen, and the transport of the conserved quantity in the boundary layer happens through fluctuations.

It might seem strange that such a large wavelength appears with the radial (wall normal) velocity near the wall, considering the impermeability condition. In channels and PC, the large scale structures reflect near the wall on only the streamwise velocity. In TC however, if present, they reflect on both the streamwise and the wall-normal velocity near the wall. This is probably due to the inherent linear instability of TC flow, which causes the formation of the rolls.

\begin{figure}
  \centering
  \includegraphics[width=0.49\textwidth]{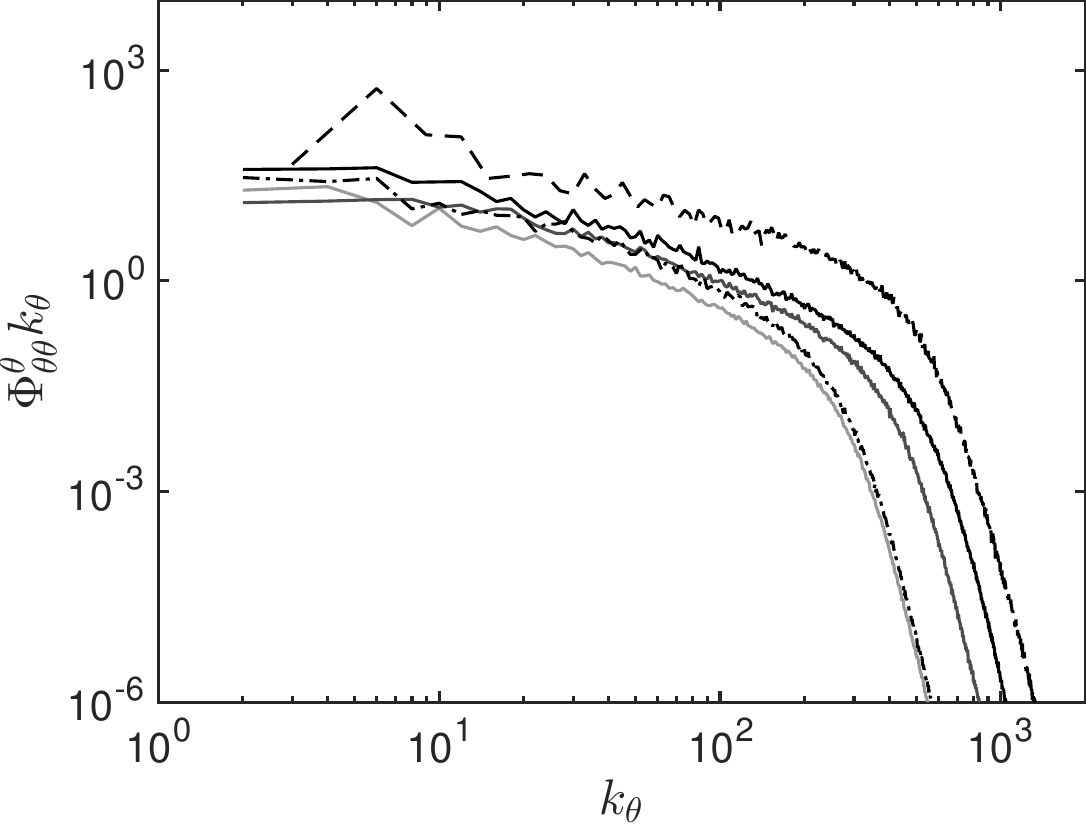}
  \includegraphics[width=0.49\textwidth]{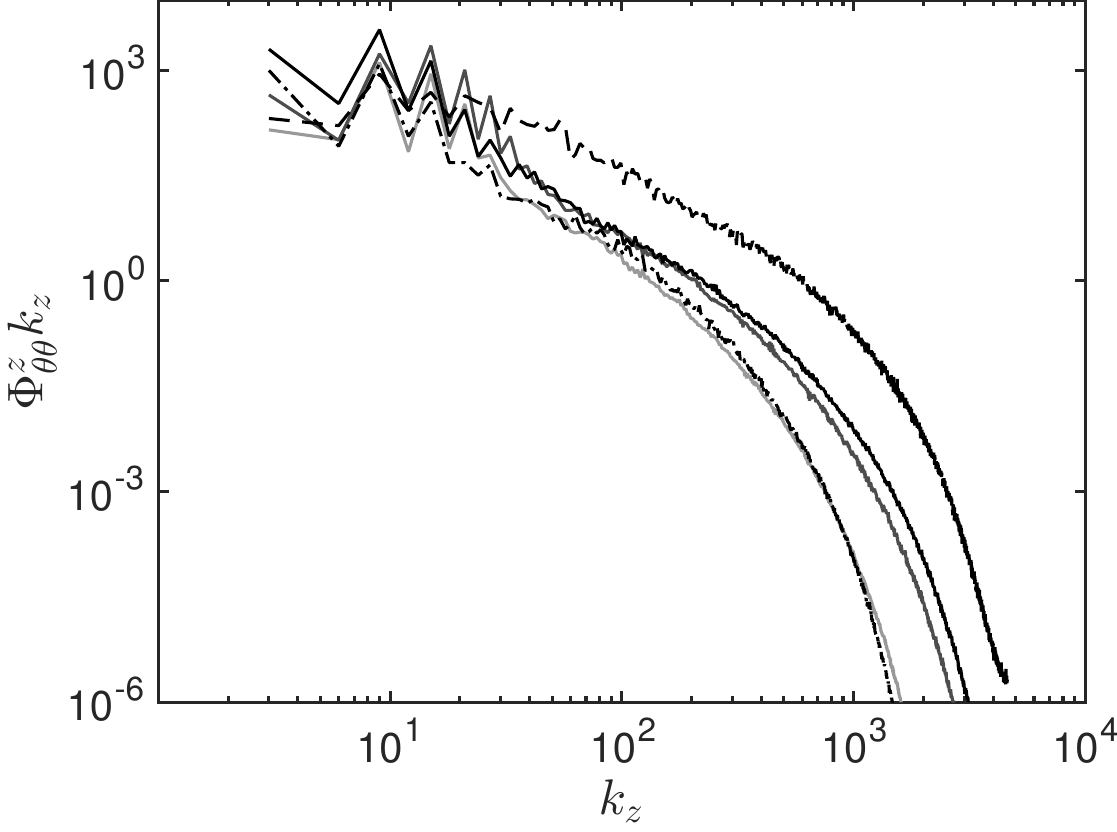}\\
  \caption{ Azimuthal and axial spectra for the azimuthal components for all simulations at mid-gap. Symbols: R1-R3 solid curves from light gray (R1) to black (R3), E5 black dashed curve, AF black dash-dot curve.  }
\label{fig:slab9spectra}
\end{figure}

Figure \ref{fig:slab9spectra} shows that rolls also dominate the spectra at the mid-gap for the small-gap cases. Once again, a prominent sawtooth behaviour in the axial spectra can be seen for the lowest wavenumbers. The spectra are consistent with the experiments by \cite{lew99}, and neither show $-1$ nor $-5/3$ scaling. This is unlike the case of pipe flow \citep{per86}, for which a $-1$ scaling was found, and also unlike the case of curved channel flow \citep{hun79}, which show the expected $-1$ scaling laws for the streamwise energy spectra in the streamwise direction. Also, the axial velocity spectra in the azimuthal direction do not show a $-5/3$ exponent, consistent with the experiments of \cite{per86}. 

When compared to other flows, it seems clear that for small gap TC, the large-scale structures are responsible for the transport of the conserved quantity. The fluctuations transport on average very little, even though their instantaneous transport can be orders of magnitude higher than the mean transport \citep{hui12}. For large gap TC flow, transport happens through the small scale fluctuations, as the rolls do not form. We note that unlike in channel flow and curved channel flow, the Reynolds stresses in TC are maximal at the mid-gap to satisfy the conservation of angular velocity current (torque). In PC, large scale structures also form, but these are also inactive, i.e. they do not transport shear \citep{avs14}. Therefore, it seems that small gap TC flow is the only flow examined up to date which involves transport by Reynolds stresses of a conserved quantity, in this case angular velocity, by large scale structures which are attached to the wall.  We note that \cite{loz12} do see large-scale wall-attached active structures in channel flow when using an analysis based on a full quantification of three-dimensional structures (Qs) instead of the more ``classic'' co-spectra. They find very little counter-gradient (i.e. negative transport) wall-attached eddies, while the Taylor roll in TC flow has both a gradient and a counter-gradient part \citep{hui12,ost14c}.  To further understand the differences, an analysis of TC using the same method is necessary, and will lead to a better one-to-one comparison. We may also speculate that a similar behaviour could also be present in Rayleigh-B\'enard (RB) flow, the flow in a fluid layer heated from below and cooled from above. RB flow also shows a linear instability \citep{cha81}, and persistent large-scale structures which cover the entire gap \citep{ahl09}.

\begin{figure}
  \centering
  \includegraphics[width=0.49\textwidth]{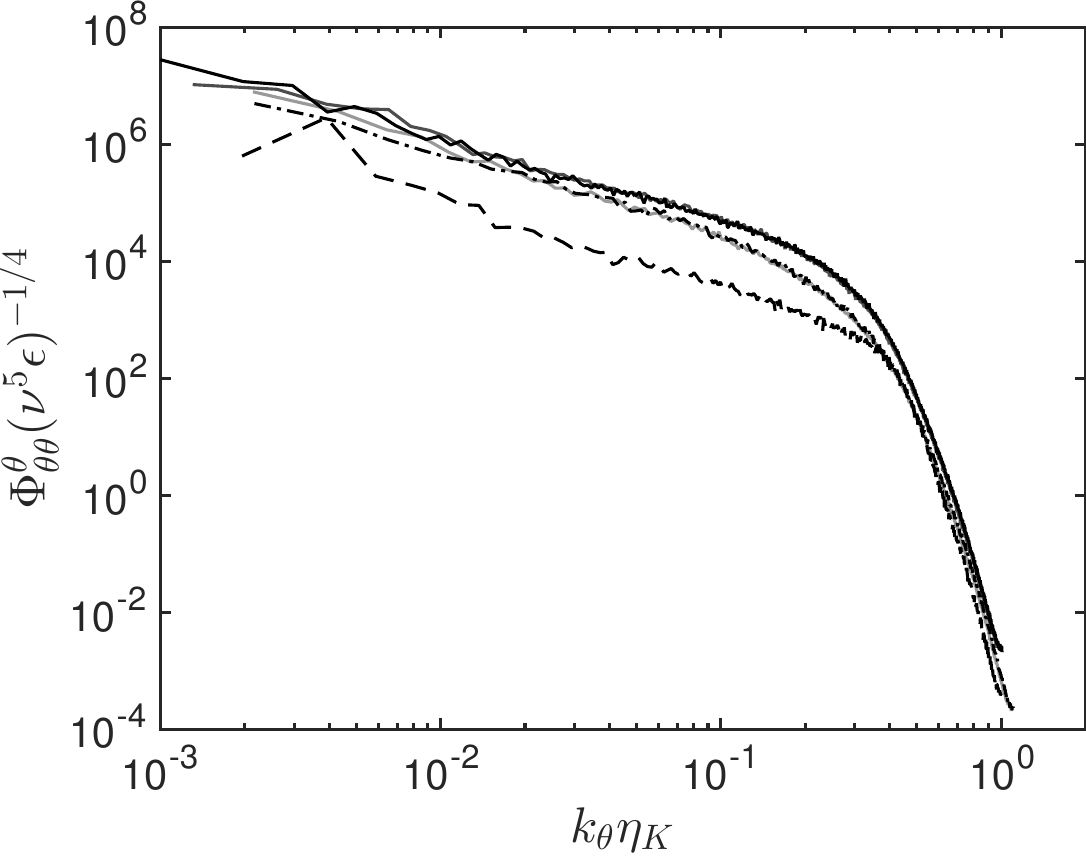}
  \includegraphics[width=0.49\textwidth]{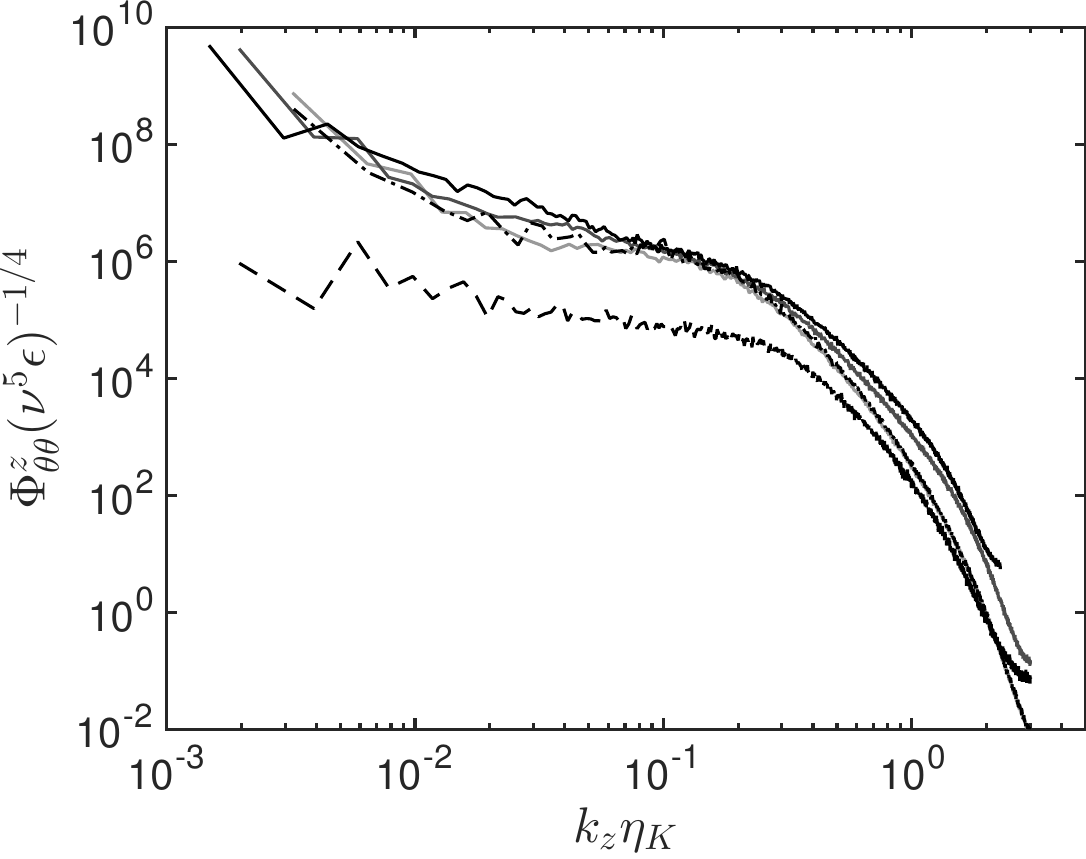}\\
  \includegraphics[width=0.49\textwidth]{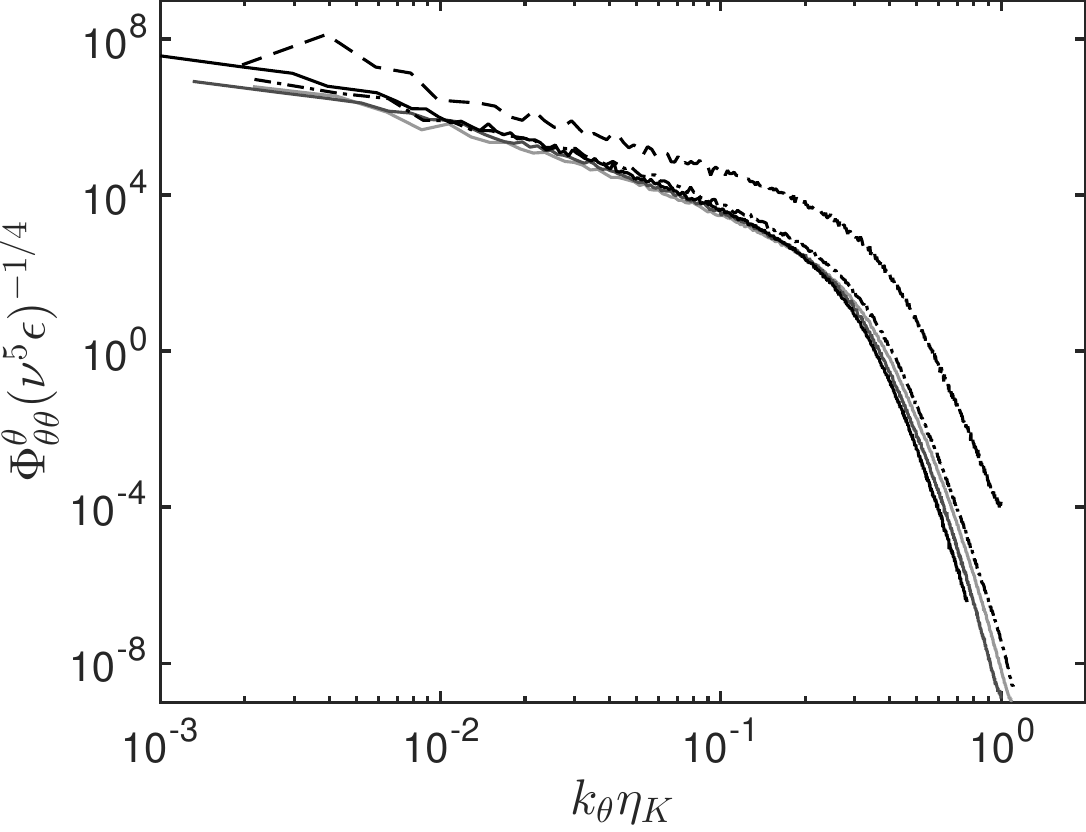}
  \includegraphics[width=0.49\textwidth]{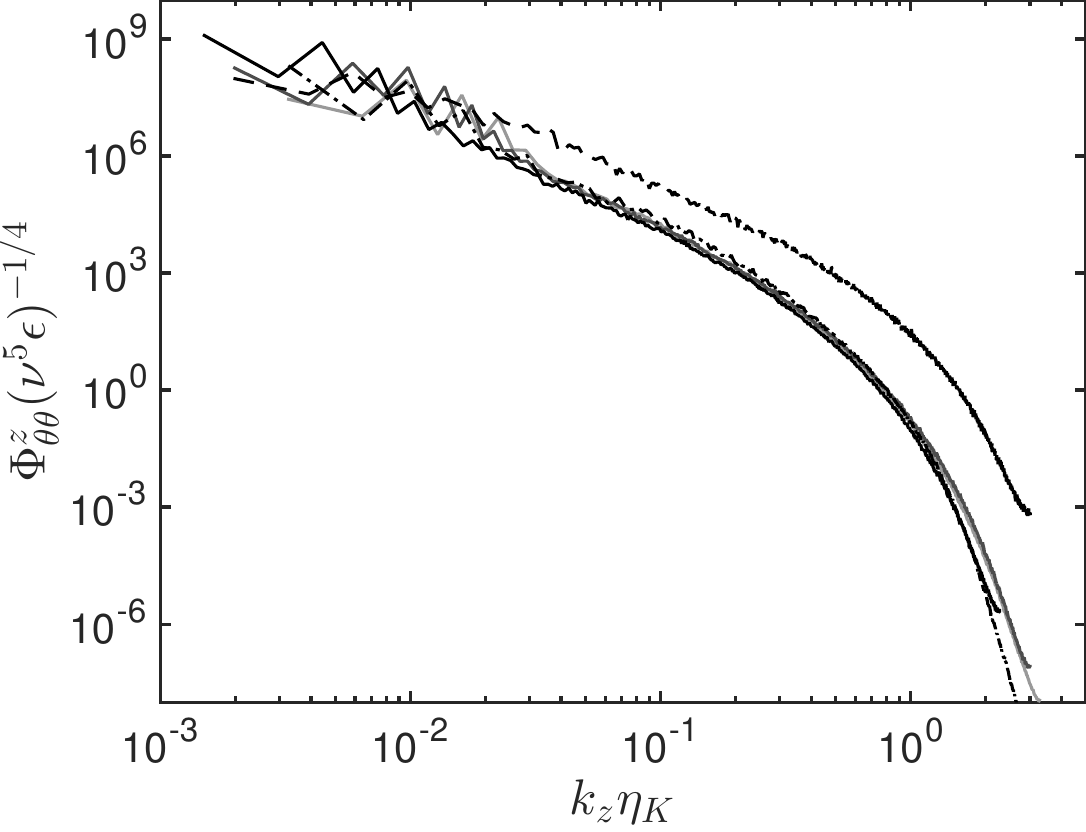}\\
  \caption{ Azimuthal (left) and axial (right) spectra for the azimuthal velocity for all simulations at $r^+\approx 12$ (top) and mid-gap (bottom). Symbols: R1-R3 solid curves from light gray (R1) to black (R3), E5 black dashed curve, AF black dash-dot curve. }
\label{fig:spectrakolm}
\end{figure}

Finally, in figure \ref{fig:spectrakolm} we show the azimuthal velocity spectra in Kolmogorov units, using the Kolmogorov length scale $\eta_K=\nu^{3/4}\epsilon^{-1/4}$, for normalization, where $\epsilon$ is the energy dissipation rate both at $r^+\approx 12$ and at the mid-gap.  At low wave-numbers the energy is dominated by the rolls, however, at high wavenumber some collapse between the small gap cases can be seen for both azimuthal spectra, and for the axial spectra at the mid-gap. All azimuthal spectra show similar shape, and the characteristic bend corresponding to $10\eta_K$, while the axial spectra can show different shapes. The most notable lack of collapse is seen for $r^+\approx 12$ in the axial spectra, but this is probably due to the different size of the herringbone streaks (which scale with $\nom$, and not $\eta_K\sim\nom^{-1/4}Re_s^{-1/2}$) or it might also be due to insufficient resolution to capture the spectra accurately for the larger runs. On the other hand, a remarkable collapse for all small-gap cases is seen for both spectra at mid-gap. Similar spectra and collapses are seen for the other two velocity components, not shown here.

\section{Summary and outlook}
\label{sec:summ}

Four large DNS simulations of small gap, and one of large gap Taylor-Couette flow were conducted, reaching frictional Reynolds numbers of $Re_\tau=4000$. In the small gap case, large scale structures, known as Taylor rolls, form in the bulk and play an active role in the system dynamics by transporting angular velocity through Reynolds stresses. The rolls are ``attached'' to the cylinders, and even deep inside in the boundary layers, for $r^+\approx 12$, their signature is significant in the $\Phi_{\theta r}$ cospectra. Adding a weak axial pressure gradient was found to advect the rolls slowly, but not to weaken them significantly, in agreement with what is seen at low Reynolds numbers. This axial pressure gradient was weak enough only to convect the rolls. Different results and flow topologies could arise from stronger gradients. The rolls signatures were still present in axial spectra in the boundary layer, and in the axial correlations of radial velocity. These structures were notably in the large gap case, where the transport of the conserved quantity is by small scale fluctuations.

The mean velocity profile of TC has quite some differences to the Prandtl-von Karman log-layer profile $U^+=\kappa^{-1}\ln y^+ + B$ with $\kappa\approx 0.4$ and $B\approx 5.2$. In the small gap case close to the walls, $\kappa$ is very close to $0.4$, while $B$ is slightly lower and Reynolds-dependent. With increasing distance from the wall, large deviations from the logarithmic law are found. The Taylor rolls redistribute angular momentum, which results in an essentially flat angular momentum profile, even for curvatures of $1\%$. This is consistent with the notion that curvature effects are orders of magnitude larger than expected by mere dimensional analysis \citep{bra73}. For the large-gap case, a degree of logarithmic behaviour can be seen for the angular velocity, consistent with the prediction of \cite{gro14}, and similar to what was already seen for $\eta=0.714$ in \cite{ost14}. This logarithmic region has a very different inverse slope $\kappa$, and only extends less than half a decade in $r^+$ for $Re_\tau\approx 3100$, due to the mixing of angular momentum in the bulk. The nature of this region seems to be very different from the classical law-of-the-wall.

Fluctuation profiles were also analysed. When adding a mean axial flow, the azimuthal velocity fluctuations presented coincide with those seen in other canonical flows at comparable $Re_\tau$. The velocity fluctuations were found to collapse when plotted in outer units, indicating the presence of an overlap region. Based on these findings, we expect that simulations at higher $Re_\tau$ will provide enough decoupling of scales for a large overlap region to form. This will allow logarithmic profiles to develop in both the mean azimuthal velocity, and the velocity and pressure fluctuations. Once again, the large-gap case was found to have a different behaviour. No rolls were present and the level of fluctuations was smaller at a comparable $Re_\tau$.

Further investigations should also look at the effect of strongly counter-rotating cylinders, which have a significant Rayleigh-stable region, or even pure outer cylinder rotation, which is completely Rayleigh-stable, and how this affects the large scale structures. Finally, additional simulations with a larger choice of pressure gradients, both in the axial direction, as done in the manuscript, as in the azimuthal direction, to provide a mixed Couette-Pouiseulle flow will be useful to increase the understanding of the system.

\emph{Acknowledgments:} We would like to thank M. Bernardini, S. Pirozzolli, E. P. van der Poel, P. Orlandi for various stimulating discussions. We acknowledge that the results of this research have been achieved using the PRACE project 2013091966 resource CURIE based in France at Genci/CEA. We also acknowledge support by an ERC advanced grant. 

\bibliographystyle{jfm}

\bibliography{literatur}

\end{document}